\documentclass{article}
\usepackage{graphicx}
\usepackage{xcolor}
\usepackage{amsfonts}
\usepackage{amssymb}
\def\ligne#1{\hbox to \hsize{#1}}
\def\leurre{\noindent\leftskip 0pt\small\baselineskip 10pt}

\newtheorem{fig}{\textbf{Figure}}
\newtheorem{tab}{\textbf{Table}}

\author{Maurice Margenstern}
\title{What can we learn from universal Turing machines?}

\begin{document}
\maketitle

\begin{abstract}
In the present paper, we construct what we call a {\bf pedagogical} universal Turing 
machine. We try to understand which comparisons with biological phenomena can be deduced
from its encoding and from its working.
\end{abstract}

\section{Introduction}\label{intro}
In one of my papers, sorry, I do not remember in which one, I wrote that if we encode
one of the smallest universal Turing machines in a DNA way, we get something whose size
is much smaller than the smallest virus. As far as a universal Turing machine has an
unpredictable behaviour, the parallel we indicated suggested that the behaviour of a
virus is more unpredictable. At that time, I did not argue more than those few words
which, clearly, did not raise any noise. As far as at the present moment people are more
concerned with what we know about viruses, it can be interesting to go back to
that comparison and to discuss whether it is relevant or not.

In Section~\ref{tiny}, we remember the results about small universal Turing machines
and classical results about Turing machines. Then, we remind the reader results about
the smallest universal Turing machines. In Section~\ref{pedago}, we introduce the 
informal notion of a {\bf pedagogical} universal Turing machine and we build such a 
machine. In Section~\ref{dispute} we discuss whether the comparison which is indicated 
in the first paragraph of the present section is relevant or not. 
In Section~\ref{conclude}, we put a temporary end to the dispute.

\section{Universal Turing machines and tiny ones}\label{tiny}

   In the present paper, the reader is supposed to know the definition of a Turing 
machine. To make things as clear as possible, we indicate that here, we only consider
deterministic Turing machines with one head and a single infinite tape in both directions.
The instructions of a Turing machine will be represented in a table: the columns are 
labelled with the alphabet of the machine, the lines are labelled with its set of
states. We remind the reader that a configuration of a Turing machine is the smallest 
finite segment of its tape containing the scanned square outside which all squares are 
empty together with the position and the state of the head. All machines we consider in
the paper have a finite initial configuration from which it clearly follows that at
any time during its computation, the configuration of a Turing machine is also finite.

\def\traitV{\vrule width 0.5pt height 10pt depth 6pt}
\def\traitH{\vrule width \hsize height 0.25pt depth 0.25pt}
\def\lignev #1 #2 #3 #4 #5 #6 {
\ligne{\hfill\hbox to 20pt{\hfill\tt #1\hfill}\traitV
\hbox to 50pt{\hfill\tt #2\hfill}\traitV
\hbox to 50pt{\hfill\tt #3\hfill}\traitV
\hbox to 50pt{\hfill\tt #4\hfill}\traitV
\hbox to 50pt{\hfill\tt #5\hfill}\traitV
\hbox to 50pt{\hfill\tt #6\hfill}\traitV
\hfill}
}

\def\smallM#1{\hbox{\small$#1$}}
\def\ftt#1{\hbox{\footnotesize\tt #1}}

As an example, we give the instructions of a Turing machine which performs an addition
over two numbers written in unary representation: $n$ is represented by $n$ vertical
strokes. The initial configuration looks like that one :
\vskip 5pt
\ligne{\hfill
\ftt{\_ * $\vert^n$ * $\vert^m$ * \_}
\hfill(1)\hfill}

In that configuration, \_ represents the symbol which means that the square of the tape
containing it is empty. It means that there is no information in the square.
A simple machine consists in replacing the {\ftt *} in the middle by a {\smallM |} 
and then erase the rightmost {\ftt *} and replacing the rightmost {\smallM |} by a 
{\ftt *}.

The machine of Table~\ref{tbaddcourt} is what I call a {\bf courteous} Turing machine.
It is defined by two conditions : the machine head never goes to the left-hand side of its
initial position and in the final configuration, the result is concatenated to the initial
data. For a courteous addition, the final configuration corresponding to (1) is:
\vskip 5pt
\ligne{\hfill
\ftt{\_ * $\vert^n$ * $\vert^m$ * $\vert^{n+m}$ * \_}
\hfill(2)\hfill}

\ligne{\hfill
\vtop{\leftskip 0pt\parindent 0pt\hsize=300pt
\begin{tab}\label{tbaddcourt}
\leurre Table of the Turing machine for the courteous addition.
\end{tab}
\lignev {} {\_} * | a X
\vskip-5pt
\ligne{\traitH}
\vskip-2pt
\lignev {1} {} {X \smallM R 2} {} {} {}
\vskip-5pt
\ligne{\traitH}
\vskip-2pt
\lignev {2} {} {\smallM R 3} {\smallM R } {} {}
\vskip-5pt
\ligne{\traitH}
\vskip-2pt
\lignev {3} {} {X \smallM L 4} {\smallM R } {} {}
\vskip-5pt
\ligne{\traitH}
\vskip-2pt
\lignev {4} {} {\smallM L } {a \smallM R 5} {} {\smallM R 7}
\vskip-5pt
\ligne{\traitH}
\vskip-2pt
\lignev {5} {| \smallM L 6} {\smallM R } {\smallM R } {} {\smallM R }
\vskip-5pt
\ligne{\traitH}
\vskip-2pt
\lignev {6} {} {\smallM L } {\smallM L } {| \smallM L 4} {\smallM L }
\vskip-5pt
\ligne{\traitH}
\vskip-2pt
\lignev {7} {} {\smallM R 8} {\smallM R } {} {}
\vskip-5pt
\ligne{\traitH}
\vskip-2pt
\lignev {8} {* \smallM L 9} {} {\smallM R } {} {* \smallM R}
\vskip-5pt
\ligne{\traitH}
\vskip-2pt
\lignev {9} {} {\smallM L} {\smallM L } {} {* \smallM !}
\vskip-5pt
\ligne{\traitH}
}
\hfill}
\vskip 10pt
   Let us explain the notation used for Table~\ref{tbaddcourt} which follows Minsky's
conventions as formulated in~\cite{minsky}. The format of an instruction is
\hbox{\tt A$M$s} where {\tt A} is the letter written by the machine in the scanned
square in place of the letter it has seen, $M$ is the move of the head: either to left,
symbolised by $L$, to right, symbolised by~$R$ or no move at all symbolised by $Z$.
When a symbol is missing, it means that it is the same as the label of the column if it is
a letter, or the same as the number of the line if it is a state or $Z$ if it is a
move. Those conventions are illustrated by the table. The symbol '!' means the halting
state: it stops the computation. Note that the instruction represented by an empty entry 
may also be considered as halting the computation: the scanned symbol is unchanged, the 
state of the head remains the same and the head scans the same square and all this 
forever.  Accordingly, when such an instruction is performed no change happens after 
that over the configuration which endlessly remains the same. We shall consider that 
situation as identical to a halting instruction as far is it can easily be detected.

   The working of the machine is rather simple: it marks the leftmost and the rightmost
{\ftt *} by {\ftt X}, states~1, 2 and 3, and then, it marks the current read $|$ by 
{\ftt a} which triggers the copying of $|$ on the leftmost empty square after the 
rightmost {\ftt X}: states~4, 5 and~6. The end of the computation is detected under 
state~4 when marking a $|$ fails as far as the rightmost {\ftt X} is met. States 7 and~8 
allow the machine to write extremal {\ftt *}'s while replacing all {\ftt X} by {\ftt *}. 
State 9 allows the machine to stop at its initial position.

   We can see that many instructions have a single symbol: that of the move. I call
such an instruction a {\bf glide}: it is particularly clear under states 5 and~6:
the machine head runs over the configuration while it meets some expected symbol in order 
to change the direction of its motion. Note that there are 13 glides in the table,
21 entries of the table are empty, meaning halting instructions as already mentioned.

\subsection{Universal Turing machine}\label{sbuniv}

   It is now time to consider universal Turing machines. In what sense such a machine is
universal? The meaning is the following: a Turing machine is called {\bf universal} 
it it is able to simulate any other Turing machine. Let us make it more precise. 
Let $M$ be a Turing machine and $C$ be an initial configuration for $M$. The 
{\bf computation} of~$M$ starting from~$C$ is a sequence \hbox{$\{C_k\}_{k\geq0}$} of 
configurations for~$M$ such that $C_0 = C$ and $C_{k+1}$ is obtained from $C_k$ by the 
application of the single instruction of~$M$ which can be applied to~$C_k$. 
The computation is a finite sequence if and only if
the last configuration in the sequence contains the halting state.
A machine~$U$ is said universal if, for any Turing 
machine $M$ and for any initial configuration~$C$ for~$M$, there is a configuration~$K$
of~$U$ such that the computation \hbox{$\{K_n\}_{n\geq0}$} of $U$ starting from~$K$ 
contains a subsequence \hbox{$\{K_{n_k}\}_{k\geq0}$} such that 
\hbox{$K_{n_k} = \tau(C_k)$} where $\tau$ is a transformation of configurations for~$M$
which does not depend neither from~$C$ nor from~$M$.

   Accordingly, a universal Turing machine $U$ must be able to simulate {\it any} Turing
machine~$M$, even if the number of states of~$M$ is bigger than that of~$U$, even if the
size of the alphabet of~$M$ is bigger than the size of the alphabet of~$U$. Turing proved
that it is possible and the key point for that is the construction of~$\tau$ and of~$K$.
A solution to that problem is to define the transformation $\tau$ as a {\bf translation} 
of the instructions of~$M$ as well as the squares of its tape. Although the tape
of a Turing machine is infinite, its configuration at any time is finite which clearly 
follows from the above definition of a configuration. The number of instructions of~$M$
also is finite so that those finitely many finite elements can be, in principle, 
translated on the tape of~$U$.

    The idea for that is to split $K$ into two parts: one contains the $\tau(I)$ for $I$
running over the elements of the table of $M$, the other contains the $\tau(x)$ for $x$
running over the configuration of~$M$ at the considered time of its computation. The place
of the head and its state must be marked in some way. If that point is satisfied, the
working of~$U$ is simple. It locates the $\tau(I)$ for the $I$ which applies to $C_k$
for the considered time~$k$. It copies $\tau(y)$ where $y$ is the letter in~$I$ onto
$\tau(\xi)$ where $\xi$ is the square of~$C_k$ scanned by the head of~$M$. When it is
performed, $U$ moves the position of the head of~$M$ from where it is in $C_k$ onto
its place in $C_{k+1}$ under construction. When the head is in its new place, $U$ changes
the state of the head of~$M$ to the state indicated by $\tau(I)$. In Section~\ref{pedago}
we describe an explicit universal Turing machine performing the just described behaviour.
We will bring in a tuning of a few points which will be explained in that section.

    How big is such a universal Turing machine? Roughly speaking, less than twenty letters
and less than a hundred of states are enough to make the table of a universal Turing 
machine. Now, if we compose a universal Turing machine~$U$ with a machine which does 
nothing, we obtain a new universal Turing machine whose table strictly contains that 
of~$U$. Accordingly, there are infinitely many universal Turing machines. Note that it 
is easy to make a bigger universal Turing machine compared with another given one. Is 
it possible to make a smaller universal Turing machine?

    The answer is yes, if we start from a rather big universal Turing machine.

    Small universal Turing machines happened to be a source of important works. It is 
not the place in this paper to report the history of that race. We refer to~\cite{mmMCU}
for such a sketchy account and to~\cite{nearywoodsa} were a more recent state of the art
is described. The key reduction for the table of a universal Turing machine is to use
the {\it delayed} computation of a Turing machine. Instead of directly simulating a Turing
machine, we allow the machine supposed to be universal to simulate another way of 
computation which, in its turn, is able to simulate any Turing machine. Such a system
is used in Rogozhin's paper~\cite{rogozhina} which gave a decisive impulse to the race
to universal Turing machines as small as possible.

\subsection{Tiny universal Turing machines}\label{sbtiny}

The results of~\cite{rogozhina} were a long time the best results. Initially appearing 
in a Soviet journal, \cite{rogozhinb} was an improved presentation of the results in
{\it Theoretical Computer Science}, far much accessible. More than twenty 
years later, Neary and Woods obtained smaller machines, see \cite{nearywoodsb}. As far as 
the present paper does not aim at giving an account of that race, we will present a 
small universal Turing machine contained in~\cite{nearywoodsb}, the machine with four 
symbols and six states. That machine improves the machine with four symbols and seven 
states of~\cite{rogozhinb}. That machine, as well as all machines of~\cite{rogozhina}
and~\cite{nearywoodsb}, is universal in the following meaning: it simulates another 
system of computation which is able to simulate any Turing machine. The program of the 
machine is reproduced by Table~\ref{tbneary} under a translation of the alphabet
to which we turn back in Section~\ref{dispute}.

\def\traitV{\vrule width 0.5pt height 10pt depth 6pt}
\def\traitH{\vrule width \hsize height 0.25pt depth 0.25pt}
\def\ligneiv #1 #2 #3 #4 #5 {
\ligne{\hfill\hbox to 20pt{\hfill\tt #1\hfill}\traitV
\hbox to 50pt{\hfill\tt #2\hfill}\traitV
\hbox to 50pt{\hfill\tt #3\hfill}\traitV
\hbox to 50pt{\hfill\tt #4\hfill}\traitV
\hbox to 50pt{\hfill\tt #5\hfill}\traitV\hfill}
}

\ligne{\hfill
\vtop{\leftskip 0pt\parindent 0pt\hsize=250pt
\begin{tab}\label{tbneary}
\leurre Table of the machine with four letters and six states by T. Neary and D. Woods.
\end{tab}
\ligneiv {} A C G U  
\vskip-5pt
\ligne{\traitH}
\vskip-2pt
\ligneiv {1} {U \smallM L } {G \smallM L } {C \smallM L } {A \smallM R 2}
\vskip-5pt
\ligne{\traitH}
\vskip-2pt
\ligneiv {2} {G \smallM R 5} {G \smallM R } {\smallM R 1} {A \smallM R }
\vskip-5pt
\ligne{\traitH}
\vskip-2pt
\ligneiv {3} {U \smallM L } {\smallM L 5} {C \smallM L } {\smallM L 5}
\vskip-5pt
\ligne{\traitH}
\vskip-2pt
\ligneiv {4} {\smallM R 5} {G \smallM R } {C \smallM R 2} {A \smallM R}
\vskip-5pt
\ligne{\traitH}
\vskip-2pt
\ligneiv {5} {C \smallM L 3} {G \smallM R 6} {C \smallM L 6} {\smallM R }
\vskip-5pt
\ligne{\traitH}
\vskip-2pt
\ligneiv {6} {} {G \smallM R 5} {C \smallM L 4} {G \smallM R 1}
\vskip-5pt
\ligne{\traitH}
}
\hfill}
\vskip 10pt

The table displays those conventions. As examples, we have the instruction when reading 
{\tt A} in state~{\tt 1}: \hbox{\tt U $L$}, and also the instruction when the head reads
{\tt U} in state~{\tt 6}: $R$. We call that latter move a {\bf glide} as far as the head
passes over the symbol without changing it and remaining in the same state.
From the table, we can see that the machine has 23 instructions. The missing instruction
for the entry for the letter {\tt A} and state~{\tt 6} is in fact the instruction
\hbox{\tt A$S$1}. Accordingly, that instruction makes the head stay over the same
square without changing its content. We already mentioned that we consider that it is a 
halting instruction.
The writing of the table requires at least 54 letters: those used in the writing of the
instructions of Table~\ref{tbneary}. Without Minsky's conventions, we need 72 letters.
In that counting, each letter of the alphabet of the machine, each symbol of move
and each number for a state counts for one unit. From our discussion about a universal 
Turing machine, we can see that the number of letters depends on the encoding we use.
As an example, if we used a binary encoding, the four letters of the alphabet require
two bits with the convention that {\tt A}, {\tt C}, {\tt G}, {\tt U} are encoded by
\ftt{00}, \ftt{01}, \ftt{10}, \ftt{11} respectively. We can encode $L$, $R$ and $S$
by \ftt{01}, \ftt{10}, \ftt{00} respectively. We can encode the number of a state
by its binary representation. If we wish to encode the table itself, we need a delimiter
for the instructions and a delimiter for the states. We deal with that question in
Section~\ref{pedago} to which we now turn.

\section{Pedagogical universal Turing machines}\label{pedago}

As indicated in the name {\bf pedagogical universal Turing machine}, the machines we 
consider under that terminology defines something which should be easily understood. 
That very
condition makes it impossible to provide a formal definition for that notion: what is 
indeed easily understandable? What another person understands can be not understood by
me and, sometimes, conversely. So that such a notion is clearly subjective. However,
I think that if somebody actually knows and understands what a universal Turing machine 
is, that person can understand the working of the pedagogical machine we give in the 
present section.

\subsection{Working of the pedagogical universal machine}\label{sbpedawork}

   As mentioned in Section~\ref{tiny}, we consider a deterministic Turing machine with
a single tape and a single head. In Sub section~\ref{sbuniv} we stressed the feature
that a universal Turing machine~$U$ must be able to simulate any Turing machine~$M$, 
whatever the size of the alphabet~$A$ of~$M$ and whatever the number~$N$ of states of~$M$.
The solution to those constraints is to encode the letters of~$A$ as well as the numbers
in $\{1..N\}$. Usually, the encoding is conceived in order to facilitate the location
of instructions and of the new state. The most convenient solution is to represent the
letters by their rank in a fixed ordering of~$A$ and to do the same for the states.
Then, the location is easy: it is enough to mark the instructions by a delimiter, to
gather in the same area the instructions depending on the same current state of~$M$
and to delimit also such an area. Then the location is obtained by a one-to-one 
correspondence between the number of symbols in the unary representation of the number
of a letter or of a state and the number of delimiters to cross in order to access
the needed area. Basically, the working is the same as what the courteous addition 
performed in Section~\ref{tiny}.

    When the needed instruction is obtained, the replacement of the content of the 
scanned square by the letter given by the instruction is performed by a copying process. 
The unary representation makes that operation somewhat complex: if the length of the
new letter is equal to that of the scanned one, there is a simple copying process to
perform. Otherwise, the situation is more complex if the length of the new letter
is different from that of the scanned one. If it is shorter, the square has to be shrunk 
which triggers to move the rest of the tape to left. If it is larger, the square has to 
be widened which entails to move the rest of the tape to right.

    The move of the whole tape of the simulated machine is also needed if the head goes 
to the left-hand side of the leftmost square of the simulated configuration. Fortunately,
we may assume that our universal Turing machine has to only simulate Turing machines
which work on a half-tape only, {\it i.e.} the tape is infinite to the right only.
We say that such a machine which observes one condition of courtesy is {\bf polite}.
Indeed such machines can simulate any Turing machine~$M$ on a bi-infinite tape: it is 
enough to imagine that the half tape is divided into two parts : one devoted to the 
right-hand side half of the tape of~$M$ and the other part is devoted to its left-hand 
side part. That entails a larger alphabet and a bigger number of states only. It is 
easily feasible. Another constraint, in order to make the code lighter is to forbid the
stationary state: the head has always to move, either to left or to right, otherwise
the instruction is considered as a halting one. That constraint does not alter the 
generality of the result as far as a machine with stationary instructions can be 
simulated by a machine which rules out such instructions. The price to pay is to allow
the machine to use more states.

    The price to pay with unary representation of the numbers is a longer representation 
of each element in the code of the universal machine. In order to get a shorter code for 
our pedagogical universal machine, we decided to represent the encoded numbers in binary.
The counterpart is a complexification of the location process. In a first step, $U$, our
universal machine, transforms the binary representation into a unary one. In a second 
step, the temporary unary representation is used to locate the expected element. 

    In order to evaluate the maximal size needed for such a transformation, $U$ counts
the delimiters for the state areas and also those for the instructions in such an area,
keeping the largest one written in unary. In Sub-section~\ref{sbpedabuild}, we more
precisely describe the process together with its implementation in the code of~$U$.
Roughly speaking, we implement the function \hbox{$n\mapsto 2^n$} together with the 
reverse function.

\def\UU{\hbox{\footnotesize\tt U}}
\def\WW{\hbox{\footnotesize\tt W}}
\def\HH{\hbox{\footnotesize\tt h}}
\def\FF{\hbox{\footnotesize\tt F}}
    The tape of~$U$ is divided into two parts: to the left-hand side $\mathcal P$,
the set of instructions collected state by state, to the right-hand side, $\mathcal T$,
an encoding of the squares of the current configuration of~$M$.
 
    The working of~$U$ can be divided into cycles where each such cycle simulates the
execution by~$M$ of one step of its computation on its tape. A cycle is divided into
five steps. At the beginning of the cycle, $U$ knows the current state of~$M$ and 
it knows which square of the tape of~$M$ is currently scanned. The area and the scanned
square are both marked by the same symbol~\WW{} which replaces the corresponding 
delimiter. The first step for~$U$ consists in transforming the binary representation of
the scanned letter in the scanned square into its unary representation stored in an 
appropriate area~$\mathcal B$ to the left-hand side of~$\mathcal P$. The second step 
consists in locating the execution of~$M$ to be performed by~$U$: $U$ counts the number 
of appropriate delimiters in $\mathcal P$ thanks to the unary representation stored 
in~$\mathcal B$. In the third step, $U$ copies the letter indicated by the 
instruction~$\mathcal I$ onto the letter of the scanned square, in $\mathcal T$. As we 
use the binary representation of numbers and we know the size of~$A$, we decide that the 
size of binary representations is standardised into a fixed size possibly using an 
additional padding symbol: if $A$ contains $n$ letters, the maximal size is 
$\vert n\vert$, the size of its binary representation, so that if $k<n$, it is 
represented by \hbox{$k_2\HH^{\vert n\vert - \vert k\vert+1}$}, where \HH{} is the 
padding symbol, and $k_2$ is the binary representation of~$k$ and $\vert k\vert$ is the 
size of $k_2$.
The third step consists in moving the head of~$M$ which is dictated by the appropriate
symbol in~$\mathcal I$. The step consists in moving~\WW{} on the previous or on the next
delimiter.
The fourth step consists in transforming the binary representation of the new state
in~$\mathcal I$ into its unary representation. The fifth step locates the delimiter of
the area devoted to the new state thanks to that unary representation: accordingly,
we are in the situation of the starting point of the next cycle.

We implement those indications in the next Sub section, completing the just given 
description by details implied by the implementation.

\subsection{Constructing a pedagogical universal machine}\label{sbpedabuild}

\def\XX{\hbox{\footnotesize\tt X}}
\def\YY{\hbox{\footnotesize\tt Y}}
\def\SS{\hbox{\footnotesize\tt S}}
\def\DD{\hbox{\footnotesize\tt M}}
\def\LL{\hbox{\footnotesize\tt L}}
\def\RR{\hbox{\footnotesize\tt R}}
\def\BB{\hbox{\footnotesize\tt \_}}
\def\hh{\hbox{\footnotesize\tt h}}
\def\zz{\hbox{\footnotesize\tt 0}}
\def\un{\hbox{\footnotesize\tt 1}}
\def\FF{\hbox{\footnotesize\tt F}}
As mentioned in Sub section~\ref{sbuniv}, the tape of~$U$ contains an encoding of
the table of~$M$ together with an encoding of the tape of~$M$. That latter 
representation is possible as far as $M$ starts from a finite configuration and as far
as at each step of its computation, only finitely many squares of its tape are non-blank
as already mentioned.

\def\PMTU{pedagogical universal Turing machine}
Figure~\ref{fgenePMTU} illustrates the basic structure
observed by the configuration of the tape of the \PMTU{} $U$. The part of the tape to the 
left hand side of the leftmost \SS{} is devoted to auxiliary computations we explain a
bit later. In between both \SS{} the tape contains an encoding of the table of~$M$.
The part of the tape to the right hand-side of the rightmost \SS{} contains an encoding
of the configuration of the tape of~$M$ restricted to a segment outside which all squares
of the tape of~$M$ are empty such that the segment also contains the head of~$M$.

\vtop{
\ligne{\hfill
\_\_{\tt F}\_ \_\_{\tt S}\_ \_\_\_{\tt S} \_\_\_
\hfill}
\begin{fig}\label{fgenePMTU}
\leurre
Basic structure of the configuration of the tape of the \PMTU.
\end{fig}	
}

The code of the program of~$M$ is a concatenation of the codes of its instructions,
provided that for each state, an instruction is present for each letter. The
code of an instruction obeys the following format:

\ligne{\hfill
\tt \YY a$_1$..$a_k$h$^\ell$\DD a$_0$..a$_m$
\hfill\rm (3)\hskip 15pt}

\noindent
where {\tt a$_i$ $\in$ $\{0,1\}$}, and \hbox{\tt a$_k$ = a$_m$ = $1$}. Indeed,
$c_\ell = \displaystyle{\sum\limits_{i=1}^k \hbox{\tt a}_i}$ and
$c_s = \displaystyle{\sum\limits_{i=0}^m \hbox{\tt a}_i}$ are numbers: 
\hbox{$c_\ell\in\{1..L\}$} and \hbox{$c_s\in\{1..N\}$} where $L$ is the number of letters
in the alphabet of~$M$ and $N$ is the number of its states. Accordingly, letters and 
states
are designated by their rank in an ordered representation of the alphabet and in
the list of states. Moreover, in order to facilitate the working of~$U$,
in (3), we assume that $k$+$\ell$ is a constant value with $\ell\geq 1$. The codes in (3)
are binary representations of the codes written with the low powers to the left.

The area which lies to the left hand-side of the leftmost \SS{} is devoted to the
computation of the values of \hbox{$n\mapsto 2^n$} for \hbox{$2^n\leq p$}, where $p$ is
the maximum between $L$ and $N$.

Let us clarify that point. The initial configuration of the tape of~$U$ is the
following one :

\ligne{\hfill
\_ \_\_{\tt S}\_ \_\_\_{\tt S} \_\_\_
\hfill\rm(4)\hskip 15pt}

The program and the tape of~$M$ are delimited to left
by~\SS{}. Each square of~$M$ is delimited to left by~\UU. The instructions of~$M$
are delimited to left by~\YY{} and the area containing the instructions attached to
a given state are delimited to left by~\XX. The tape of~$M$ is delimited to right
by~\BB{}$\,$, the blank of~$U$. The blank of~$M$ is, by convention, the first letter 
of~$A$. The area devoted to the computation of \hbox{$n\mapsto 2^n$} is delimited
to right by the leftmost~\SS{} and, to left by \BB$\,$. That area contains two
sub areas marked by \FF$\,$. In between \SS{} and \FF{}, we have in unary a 
number~$P = \max(L,N)$.
To the left-hand side of \FF, we have the representation in unary of the powers of~2
which are not greater than~$P$. We can see that area of powers of~2 as a pattern
to construct a number in unary knowing its binary representation.

As an example, the following configurations show us how to compute 111111 from 011, as 
illustrated in (5):

\ligne{\hfill
$\vcenter{\vtop{\leftskip 0pt\parindent 0pt\hsize = 200pt
\ligne{\hfill \ftt{d000d0ddF000000000S} \hfill$(a)$\hskip 15pt}
\ligne{\hfill \ftt{d000e0ehF000000000S} \hfill$(b)$\hskip 15pt}
\ligne{\hfill \ftt{d000L0ehF000000000S} \hfill$(c)$\hskip 15pt}
\ligne{\hfill \ftt{d000d0LhFhhhh00000S} \hfill$(d)$\hskip 15pt}
\ligne{\hfill \ftt{d000d0ddFhhhhhh000S} \hfill$(e)$\hskip 15pt}
}}$\hfill(5)\hskip 15pt}
\vskip 10pt
In $(a)$ we have the initial setting when it is installed by the first operation 
performed by~$U$ before the simulation itself. In $(b)$ we have the copy of 011 on the
pattern: note that the writing of the digits is on the reverse order with respect to
the source. In $(c)$ \LL{} marks the power of~2 to be copied over the 0's between \FF{}
and \SS{}. In $(d)$ the copy is performed: four \HH's replaced four {\tt 0}'s. Moreover,
the next power of~2 to be copied is marked with \LL. In $(e)$, we can see that the
copy of the last power of~2 to be copied is performed. The pattern is completely restored
and we have six \HH's which is the writing in unary of 011. Table~\ref{tbextend} is
extracted from the table of the \PMTU. It contains the instructions which perform the
transformations sketchily indicated in (5).

\def\dd{\hbox{\tt d}}
\def\ee{\hbox{\tt e}}
\def\lesvii #1 #2 #3 #4 #5 #6 #7 #8{
\hbox to 18pt{\hfill\footnotesize\tt #1\hfill}\traitV
\hbox to 26pt{\hfill\footnotesize #2\hfill}\traitV
\hbox to 26pt{\hfill\footnotesize #3\hfill}\traitV
\hbox to 26pt{\hfill\footnotesize #4\hfill}\traitV
\hbox to 26pt{\hfill\footnotesize #5\hfill}\traitV
\hbox to 26pt{\hfill\footnotesize #6\hfill}\traitV
\hbox to 26pt{\hfill\footnotesize #7\hfill}\traitV
\hbox to 26pt{\hfill\footnotesize #8\hfill}\traitV
}
\def\addauxvii #1 #2 #3{
\hbox to 26pt{\hfill\footnotesize #1\hfill}\traitV
\hbox to 26pt{\hfill\footnotesize #2\hfill}\traitV
\hbox to 26pt{\hfill\footnotesize #3\hfill}\traitV
}
\ligne{\hfill
\vtop{\leftskip 0pt\parindent 0pt \hsize=310pt
\begin{tab}\label{tbextend}
\leurre
Part of the table of the \PMTU{} which computes $n$ in unary from its
binary representation in the pattern at the left-hand side of \FF.
\end{tab}
\ligne{\hfill\lesvii {} {\zz} {\un} {\LL} {\RR} {\YY} {\UU} {\FF}
\addauxvii h d e \hfill}
\vskip-6pt
\traitH
\vskip-2pt
\ligne{\hfill\lesvii {40} {\RR} {} {} {} {} {\RR} {{\tt R41}}
\addauxvii {\RR} {h \RR} {\RR} \hfill}
\vskip-6pt
\traitH
\vskip-2pt
\ligne{\hfill\lesvii {41} {{\tt hL42}} {\LL} {} {} {} {} {}
\addauxvii {\RR} {} {{\tt L42}} \hfill}
\vskip-6pt
\traitH
\vskip-2pt
\ligne{\hfill\lesvii {42} {\LL} {\LL} {{\tt R43}} {\LL} {\LL} {\LL} {\LL}
\addauxvii {\LL} {\LL} {\LL} \hfill}
\vskip-6pt
\traitH
\vskip-2pt
\ligne{\hfill\lesvii {43} {{\tt YR40}} {} {} {\RR} {\RR} {\RR} {{\tt L44}}
\addauxvii {{\tt UR40}} {} {{\tt RR40}} \hfill}
\vskip-6pt
\traitH
\vskip-2pt
\ligne{\hfill\lesvii {44} {} {} {{\tt dR45}} {{\tt eL}} {{\tt 0L}} {{\tt dL}} {}
\addauxvii {} {} {} \hfill}
\vskip-6pt
\traitH
\vskip-2pt
\ligne{\hfill\lesvii {45} {\RR} {} {} {} {} {} {\RR}
\addauxvii {{\tt 0R46}} {\RR} {{\tt LR40}} \hfill}
\vskip-6pt
\traitH
}
\hfill}
\vskip 10pt
The reader is invited to look at the configurations of~(5) in order to better see the
transformations given in what follows. Under state~{\tt 40}, a \dd{} is transformed 
into \hh{} and the head crosses \FF{} which makes it to change the sate to~{\tt 41}.
Under that state, meeting the leftmost \zz{} in between \FF{} and~\SS, the head transforms
\zz{} into \hh{} and change its state to~{\tt 42}. State~{\tt 42} is a glide to left until
\LL{} is met which triggers the copy of the corresponding power of~2. The meeting of~\LL{}
makes the head change its motion to right and to change its state to~{\tt 43}. That state
is a glide over the symbols which are marked as copied. The glide occurs when a
not yet copied symbol is met: \zz{}, \hh{} or \ee{} which are changed to \YY{}, \UU{}
or \RR{} respectively. When a symbol is met, the head goes again to right under
state~{\tt 40}. The head glides over \UU, \hh{} and \ee, it changes \dd{} to~\hh{}
and changes its state to~{\tt 41} going on to right, running over \hh's until \zz{} is
met. That writing triggers a new cycle of copying that power of~2. That copying is 
completed when \FF{} is met under state~{\tt 43} which, looking after a symbol to copy,
meets \FF. Then the head turns to state~{\tt 44} which unmarks the copied symbol and
looks after \LL{} which is changed back to \dd{} and makes the head to turn to 
state~{\tt 45} and to go to right. Under that state, the head looks after the next \ee{}
which is changed to~\LL. While going to the right, the head restores the \zz's to the 
left-hand side of the new \LL{} which were changed to~\hh. When \ee{} is met, the head
is now under state~{\tt 40} and a new cycle of copying the current power of~2 is 
triggered. Under state~{\tt 45} too, if looking after~\ee{} the head meets \FF, it
goes to right and, meeting \hh, it knows that the transformation from binary to unary is
performed. The head turns to state~{\tt 46} which starts the process of detecting the
needed instruction. The process is already started by state~{\tt 45} which transforms
the met \hh{} to~\zz.

   That process is used twice in a cycle of computation of~$U$ devoted to the execution
of one step of~$M$. In fact, the same states are used for the conversion of the binary
representation copied on the pattern to the left-hand side of~\FF. We shall explain
that point a bit later.

   We give the full table of~$U$ in Appendix~1 to the paper, see Table~\ref{appPMTU}. 
Before going on the explanation of the precise working of~$U$, it is time to indicate 
how the tape of~$M$ is encoded. The implementation of a square of~$M$ is given by~(6):
\vskip 5pt
\ligne{\hfill
\tt \UU a$_1$..$a_k$h$^\ell$
\hfill(6)\hskip 15pt}

\noindent
where {\tt a}$_i$ is in \hbox{$\{0,1\}$}, \hbox{$\ell\geq1$} and $h+k$ is the same 
constant value as in~(3) for the instructions. That makes easier the copying of the 
letter of the instruction onto the scanned square: it overwrites the content of the
square without any comparison.

\def\aa{\hbox{\footnotesize\tt a}}
\def\TT{\hbox{\footnotesize\tt T}}
\def\ZZ{\hbox{\footnotesize\tt Z}}
   The first transformation from binary to unary is performed for the letter~$n$ 
written in the scanned square. We know that $n$ is the binary representation of the
rank of the considered letter in the alphabet of~$M$. From~(6), we know that
$n=\displaystyle{\sum\limits_{i=1}^k \aa_i2^{i-1}}$, so that the binary representation
of~$n$ on the square has the low powers of~2 to left. We note that (5) shows us that 
the binary representation over the pattern has the high powers to left. The display
is chosen in order to take benefit of the direction of the motion of the head in order
to facilitate the operations.

The alphabet of~$U$ consists of
the following symbols: 
\vskip 5pt
\ligne{\hfill
\BB, \zz, \un, \LL, \RR, \XX, \YY, \UU, \SS, \hh, \dd, \ee, \FF,
\ZZ, \TT. \hfill}
\vskip 5pt
\noindent
Note that \BB{} is the blank, {\it i.e.} the symbol indicating that the
square which contains it is empty. Symbols \LL{} and \RR{} occur in the instructions
as indication of the direction of the move to be performed by the head when it executes
the corresponding instruction while \ZZ{} indicates a halting instruction with no
mention of a direction. The symbols \XX, \YY{} and \UU{} are delimiters.
We already met \YY{} and \UU, delimiters of an instruction and of a square of the tape
respectively. The symbol \XX{} delimits the states: in between two consecutive \XX, we 
have, on the tape of~$U$, all the instructions corresponding to the left-hand side \XX.
It is the same for the last state whose instructions are displayed between the rightmost
\XX{} and the rightmost \SS. The symbol \hh{} is basically a padding symbol in the binary
representations used in instructions and squares of the tape. It is also used for marking 
other symbols during some operations. We already met the symbol~\FF. For what are
symbols \dd{} and \ee, they are auxiliary symbols used during copying processes.
Two \WW's occur on the tape: one usually replaces the \YY{} which delimits the instruction
currently executed. That \WW{} replaces the leftmost \XX{} in the initial configuration.
The other~\WW{} replaces the \UU{} which delimits the scanned square.

   The first operation performed by~$U$ is to compute the number of states and of
letters of~$M$: it is enough to compute from \SS{} the number of \YY's from the 
leftmost~\XX{} to the next one and then, again starting from \SS, to compute
the number of \XX's until the next \SS{} is met. That computation is performed by 
states~{\tt 1} up to~{\tt 11},see Sub-table~{\bf\ref{appPMTU}.a}: states from~{\tt 1} 
up to~{\tt 6} compute the indicated
\YY's and states from~{\tt 7} up to~{\tt 11} compute the \XX's. State~{\tt 11} puts
the \FF{} which closes the zone of \un's giving $P$ in unary. In that zone, the machine 
computes in unary the powers of~2 which are not greater than~$P$. That is performed by
states~{\tt 12} up to~{\tt 22}, see Sub-table~{\bf\ref{appPMTU}.b}. The first four states 
writes {\tt ddhd} to the right-hand side of~\FF. Then state~{\tt 17} looks after the 
leftmost \un{} which is transformed into~\hh. The head turns to state~{\tt 18} and 
moves to left to mark the symbol corresponding to the transformed \un: \dd{} is turned 
to~\ee{} and \hh{} in between \FF{} and the rightmost \dd{} or \ee{} is changed to~\RR. 
Under state~{\tt 18} the head goes back to left under state~{\tt 16} in order to look 
after a fresh symbol \hh{} or \dd{} to copy, marking it as \RR{} or \ee{} respectively. 
The cycle of the succession of states~{\tt 17}, {\tt 18} and {\tt 16} is repeated until 
\FF{} is met under state {\tt 16}. It means that the number of $h$'s written by 
state~{\tt 17} is equal to the number of symbols \ee{} and \RR{} in between \FF{} and 
the leftmost \hh{}. The head goes back to right under state~{\tt 19} in order to change 
the rightmost \hh{} written by state~{\tt 17} to~\dd{}, which is performed by 
state~{\tt 20} and state~{\tt 16} is called for a new cycle of copying. When \SS{} is 
reached under state~{\tt 17}, $U$ knows that the last marked power of~2 not greater 
than~$P$ is the highest one. Accordingly, that part of the computation is over. 
State~{\tt 21} is called in a motion to left in order to rewrite \zz's as \un's and to 
return \ee's and \RR's to \dd's and \hh's respectively. The task is continued by 
state~{\tt 22}. When \FF{} is reached under that state, a new action of the \PMTU{} starts beginning under state~{\tt 23}.

 In Sub-table~{\bf\ref{appPMTU}.c}, states~{\tt 23} up to~{\tt 29} construct the pattern 
 which lies to the left-hand side of~\FF{} and which is later used to compute the unary 
 representation of a number given in binary representation. The head goes form one side 
 of~\FF{} to the other, looking after the next symbol to be copied when it is to the 
 right-hand side of~\FF{} and copying it when it is to the left-hand side. States~{\tt 23} up to~{\tt 27} perform that copying. That part of Sub-table~{\bf\ref{appPMTU}.c} is 
 displayed on the left-hand side half of the sub-table.
State~{\tt 23} performs the overwriting of the symbols to be copied, namely \zz, \hh{}
and \dd. State~{\tt 24} writes \dd{} on the rightmost \BB-square, while state~{\tt 26}
performs the same operation for~\zz. As far as those states go from the marked symbol
to the copying place, when it is under those states, the head goes to left, gliding 
over~\zz, \un, \dd{} and \FF. State~{\tt 25} operates the opposite motion, leading back
the head under state~{\tt 23} when it meets \FF. The end of the transformation happens
when looking after a symbol to be marked the head under state~{\tt 23} reads \SS.
It turns to state~{\tt 27} which transforms the initial \un's in between \FF{} and~\SS{}
into \zz's and it turn to state~{\tt 28} when it meets \FF. Two states, {\tt 28} and 
{\tt 29} are needed to go to the ~\WW{} which locates the square scanned by~$M$.
States~{\tt 28} and~{\tt 29} are displayed on the right-hand side half of 
Sub-table~{\bf\ref{appPMTU}.c}.

In Sub-table~{\bf\ref{appPMTU}.d}, states~{\tt 30} to~{\tt 40} copy~$n$, the binary 
representation of the letter scanned by the head of~$M$, onto the scale-pattern lying 
on the left-hand side of~\FF. 
State~{\tt 30} operates on the square scanned by~$M$ which is located by the appropriate
\WW. It glides over \zz's and \un's which have been changed into \dd's and \ee's 
respectively until it meet a free symbol \zz{} or \un{} which it changes as already 
mentioned. Then, state~{\tt 31} is called to start the copying of a~\zz{} while 
state~{\tt 35} is called to perform the same operation for a \un. Two states are needed 
to cross the configuration of~$U$ from the square scanned by~$M$ up to \FF. In the
area delimited by~\FF{} and the rightmost \BB{} to the right-hand side of it, 
state~{\tt 32} overwrite a \dd{} as \hh{} as far as the corresponding power of two
is marked by~\zz{} in the binary representation of~$n$. Then, by states~{\tt 33} 
and~{\tt 34}, the head goes back to the scanned square and then state~{\tt 30} is called
for marking a new symbol. Similarly, state~{\tt 35} looks after~\FF{} and then 
state~{\tt 36} overwrite the fresh~\dd{} as \ee{} as far as the corresponding power of two
is marked by~\un{} in the binary representation of~$n$. That marking is performed by
state~{\tt 36} and the return to the scanned square is again initiated by state~{\tt 33}.
That copying process is completed when under state~{\tt 30} the head meets \hh. As
mentioned in (6), there is at least one \hh{} in the code of a square of the tape of~$M$.
Then state~{\tt 37} is called and the head goes to left until it meets under 
state~{\tt 38} the rightmost \BB. Then the head goes to left under state~{\tt 39} until
it finds the leftmost~\ee. Then state~{\tt 40} is called to change that \ee{} into \LL{}
in order to start the conversion of the binary representation to a unary one.

In Sub-table~{\bf \ref{appPMTU}.e}, its left-hand side half reproduces 
Table~\ref{tbextend} in another display. That sub-table displays the states from~{\tt 40}
up to~{\tt 50}. The columns, the lines are labelled by the states, the letters 
respectively. In the sub-table, there is no line for \BB{} as far as it raises only a 
halting instruction under the considered states. We explained the process of converting 
the binary representation into the unary one. States~{\tt 46} up to~{\tt 50} are devoted 
to the location of the needed instruction.
By assumption, the \WW{} in between the two \SS's of the configuration of~$U$ replaces
the \XX{} delimiting the instructions associated to the current state of~$M$. The counting
is performed by erasing a unary symbol and then mark a \YY{} in between \WW{} and the
leftmost \XX{} or \SS{} to its right-hand side. When meeting \WW{} under state~{\tt 46},
the head turns to state~{\tt 47} looking after the closest \YY{} while gliding to right.
That \YY{} is transformed into~\FF{} which triggers state~{\tt 48} and a glide to left
back to~\SS{} where state~{\tt 49} is called. Under state~{\tt 49}, the head goes back 
to the closest \hh, gliding other \zz's. When that \hh{} is met,it is marked by \zz{}
and, turning to state~{\tt 50}, the head goes back to right to the closest~\FF{} which
marks a \YY. When that \FF{} is met, the head restores the \YY{} and it goes to 
state~{\tt 47} in order to find the next \YY{} which has to be marked. The location of the
appropriate instruction is obtained when looking after a not yet erased \hh, the head 
finds \FF{} under state~{\tt 49}. It then turns to the next stage of the simulation
under state~{\tt 51}.

In Sub-table~{\bf \ref{appPMTU}.f}, states~{\tt 51} up to~{\tt 60}, $U$ copies
the letter encoded in the instruction to be executed by~$M$ onto the square of $M$ tape
squared by~$M$. First, state~{\tt 51} goes to the right-hand side \FF{} which is
replaced by~\WW: it is the instruction to be executed. Then, the  copying processes as 
usual: the first \zz{} or \un{} met by state~{\tt 52} is replaced by \dd{} or \ee{}
respectively. The meeting of a \zz{} triggers state~{\tt 53} in order to overwrite the
first not yet overwritten symbol by~\dd{} under state~{\tt 54}: the crossing of \WW{} by
state~{\tt 53} means that the square scanned by~$M$ is reached. When the marking is 
performed, the return to the \WW{} of the instruction is obtained by state~{\tt 55}.
When it is done, state~{\tt 52} is again called so that a new cycle of copying one symbol
is performed. When \un{} is overwritten under state~{\tt 52}, that triggers state~{\tt 57}
which overwrites with \ee{} the leftmost not yet transformed symbol of the square 
scanned by~$M$. States~{\tt 57} and~{\tt 58} are parallel to states~{\tt 53} and {\tt 54}.
The return state is {\tt 59} which calls state~{\tt 60} when the \WW{} delimiting the
instruction is met: state~{\tt 52} is again called so that a new cycle of copying one
symbol is performed. The process is stopped when \hh{} is read under state~{\tt 52}:
it means that the binary representation of the new letter is completely copied onto the
square scanned by~$M$. Note that during the glide to left or to right in between 
both~\WW's glides over~\dd{} and~\ee. When looking after a new symbol to be copied,
state~{\tt 52} also glides over the \dd's and \ee's present in the letter of the 
instruction.

In Sub-table~{\bf \ref{appPMTU}.g}, a first part of the states erase the markings:
\dd's and \ee's are back turned to \zz's and \un's respectively. That operation
is performed by states~{\tt 61} in the instruction, by state~{\tt 62} which makes the 
head of~$M$ go back to the scanned square and by state~{\tt 63} which clears the
marking in the scanned square. When it is completed as far as state~{\tt 63} reads \hh,
the head of~$U$ goes back to the instruction, states~{\tt 64} and~{\tt 65}, in order
to look which move of the head of~$M$ has to be performed. It is given by the letter~\RR{}
or \LL{} of the instruction transformed into~\FF{} or~\UU{} respectively. States~{\tt 67}
up to~{\tt 70} move~\WW{} to the right place. As far as $M$ is supposed to be a polite
Turing machine, the move to left raises no problem: the required \UU-delimiter will be
found. A move to right is more complex: when going to right in the scanned square,
the head of~$U$ under state~{\tt 70} meets \UU, the encoding of a square follows to the
right-hand side of that~\UU. But such an \UU{} is missing if \WW{} was put on the 
rightmost square of the tape of~$M$. In that case, the head of~$U$ meets \BB. It is
transformed into~\UU{} and an empty square must be copied, which is performed by
states in Sub-table~{\bf \ref{appPMTU}.h}.

In Sub-table~{\bf \ref{appPMTU}.h}, states~{\tt 71} up to~{\tt 79} perform the 
construction of an empty square at the right-hand side end of the configuration of~$M$
and they also control the copying process of the new state of the the head of~$M$
onto the scale-pattern at the left-hand side of~\FF{} and its conversion in a unary
representation. The construction of an empty square is performed by states~{\tt 71} up
to~{\tt 75}. The symbols of the previously scanned square $Q$ are marked in a copying
process. As far as the size of a square of~$M$ is not known but as far as, by 
construction, all squares of~$M$ have the same size, it is enough in that copying process
to replicate any symbol of~$Q$ as a \hh. In that process, \hh's, \zz's and \un's of~$Q$
are one by one replaced by~\YY, \ee{} and~\RR{} respectively, which is performed by
state~{\tt 74}. States~{\tt 71}, {\tt 72} and~{\tt 73} initialize that process.
State~{\tt 72} copies \hh{} on \BB{} each time a symbol of~$Q$ has been marked. The
construction is done when under state~{\tt 74} the head reads \UU. Then under 
state~{\tt 75}, $U$ replaces by~\un{} the leftmost~\hh{} of the just constructed square
and still under state~{\tt 75} it erases the marks in~$Q$, returning \YY's, \ee's and 
\RR's to \hh, \zz{} and \un{} respectively.

State~{\tt 77} crosses \SS{} and then state~{\tt 78} makes the head of~$U$ go back
to left to the other~\SS{} which is changed to~\TT: the head goes to~\WW, still
marking the state of the executed instruction and turns it back to~\XX. Then, under
state~{\tt 79}, the head of~$U$ looks after \UU{} or \FF{} which is restored into
\LL{} and~\RR{} respectively and the new state is~{\tt 30}. It means that
the binary representation of a number which is to the right-hand side of that move
symbol is copied onto the scale-pattern to the left-hand side of~\FF, see
Sub-table~{\bf\ref{appPMTU}.d}. State~{\tt 39} of that table calls states~{\tt 40} of
Sub-table~{\bf\ref{appPMTU}.e} which converts that binary representation into unary.
When the conversion is completed, state~{\tt 46} reads \TT, which triggers state~{\tt 80},
the first state of Sub-table~{\bf\ref{appPMTU}.i}.

In Sub-table~{\bf \ref{appPMTU}.i}, the last sub-table of Table~\ref{appPMTU},
states~{\tt 80} up to~{\tt 86} perform the location of the new state of~$M$. The
principle is the same as that performed in Sub-table~{\bf\ref{appPMTU}.e}. The
leftmost~\XX{} is replaced by \FF{} by state~{\tt 80}, corresponding to the first marking
of a \hh{} in the unary representation of the number of the new state. Then state~{\tt 84}
makes the head of~$U$ go to left until it reaches the rightmost \BB{} to the left-hand 
side of the configuration of~$U$. Then, under state~{\tt 85}, the head marks a new \hh.
That marking changes the state to~{\tt 81} under which the head goes to right until
reaching~\TT. There the head changes to state~{\tt 82}: when meeting \FF, it changes it
back to~\XX. Then under state~{\tt 83}, it still goes to right, looking after the next
\XX, which is marked and which triggers state~{\tt 84} so that a new cycle of search 
starts. When under state~{\tt 85} the head meets no more \hh, there are only \zz's 
between \FF{} and~\TT, so that the head meets~\TT. It changes~\TT{} to~\SS{} and it goes 
to right under state~{\tt 86} which changes \WW{} to the~\YY{} it previously marked and 
then replaces~\FF{} by \WW: the new state is located. The new scanned square is also located so that meeting \SS{}
under state~{\tt 86}, the head goes to left under state~{\tt 87} until it meets
again the \WW{} in the program of~$M$. Then state~{\tt 29} is called so that a new
cycle for simulating the next step of~$M$ is starting.

   Accordingly, the \PMTU{} we devised has 87 states and 16 letters which means
1392 instructions. The encoding we described for~$M$ can also be used for~$U$.
It requires 10351 letters of that encoding. The execution of $U$ on the program
of Table~\ref{tbaddcourt} as $M$ and on the encoding of *|||*||* as its data
requires 106 steps of computation for~$M$ and 1,143,717 steps for~$U$, as indicated by
the execution of the computer program~$\mathcal P$$_1$ I devised to check the 
correctness of the \PMTU. The initial configuration of the tape of~$M$ in that of~$U$ is:
\vskip 5pt
\ligne{\hfill\small\tt SW01hhU11hhU11hhU11hhU01hhU11hhU11hhU01hh\_,\hfill}
\vskip 5pt
\noindent
and its final configuration is:
\vskip 5pt
\ligne{\hfill\small\tt 
SW01hhU11hhU11hhU11hhU01hhU11hhU11hhU01hhU11hhU11hhU11hhU11hhU11hhU01hh\_,\hfill}
\vskip 5pt
\noindent
as computed by~$\mathcal P$ which is what was expected.

\section{Comparison: relevant or not?}\label{dispute}

   Let us now turn to what is raised in the introduction. What can be the basis off a
comparison between a universal Turing machine and a virus? Of course, I do not have in 
mind computer viruses which are something different based on another behaviour of
programs which aim at replicate themselves inside as most machines as possible and
to invest each infected machine, preventing it to work. A computer virus has an 
intention, a natural virus has no will. What we call will when speaking of a natural
virus, we speak of the result of natural selection on its evolution. As an example,
if we say that a natural virus tries to adapt to its hosts, that 'trying' is indeed the
result of selection: the variant of the virus which is the least malevolent to their
hosts has the best chance to survive.

    The basis of our comparison is, I think, more profound. It is grounded on the
complexity of the things we consider and on their behaviour, mainly the working of
a universal Turing machine and the behaviour of a virus at a molecular level.
    
    The unpredictability of the behaviour of a universal Turing machine is a theorem:
it is a corollary of both the existence of universal Turing machines and the algorithmic
unsolvability of the halting problem for Turing machines from which a lot of corollaries
are derived as, for instance, the same impossibility to say whether a Turing machine
launched on a given configuration reaches another fixed in advance configuration.
Before going on the dispute we postpone to Sub-section~\ref{sbcontinue}, I give
in Sub-section~\ref{sbrnacode} the description of how we can encode the code of~$U$
in RNA-terms and how it is possible to deal with that RNA-code. 

\subsection{How to construct a universal RNA Turing machine}\label{sbrnacode}

\def\AAA{\hbox{\bf\tt A}}
\def\CCC{\hbox{\bf\tt C}}
\def\GGG{\hbox{\bf\tt G}}
\def\TTT{\hbox{\bf\tt T}}
\def\UUU{\hbox{\bf\tt U}}
\def\RNAmtu{$RNA$ universal Turing machine}
    Let us look at the size of a code of a Turing machine, using the encoding defined
for the \PMTU{} $U$ in (3) and (6), in Section~\ref{sbpedabuild}. As given in 
Table~\ref{tbneary}, the code of that tiny universal Turing machine, say $N$, 
requires 206 letters in the just mentioned encoding. The code of $U$ itself in the same 
encoding requires 10,351 letters. That encoding is based on the following alphabet: 
\hbox{$\zz,\un,\LL,\RR,\XX,\YY, \UU,\WW,\SS,\hh,\dd,\ee,\FF,\ZZ,\TT$}, considering also
the working of~$U$. The ge\-nome of a DNA-virus consists in a very long chain of 
thousands of nucleotides of four kinds: adenine, cytosine, guanine and thyomine denoted 
by \AAA, \CCC, \GGG{} and \TTT{} respectively, given in alphabetic order. In the case of
an RNA-virus, the composition of the genome is similar: we have also four kinds of
nucleotides, the same ones with the exception of thyomine which is replaced by uracil,
denoted by \UUU. We later refer to the alphabet \hbox{$\{$\AAA,\CCC,\GGG,\UUU$\}$} as 
the {\bf RNA-alphabet}. 

With those four letters, we may encode the alphabet of~$U$. For example we can define
the following correspondence, encoding a letter of~$U$ by two letters of the RNA-alphabet:
\vskip 5pt
\def\www{\hskip 10.625pt}
\ligne{\hfill
$\vcenter{\vtop{\leftskip 0pt\parindent 0pt\hsize=280pt
\ligne{\hfill \tt S\www X\www Y\www U\www T\www F\www W\www \_\www L\www 
R\www 0\www 1\www h\www d\www e\www Z\hfill}
\ligne{\hfill
\tt UU  AA  CC  GG  UA  UC  UG  CG  AC  AG  CA  CU  GA  GC  GU  AU\hfill}
}}$
\hfill(7)\hskip 15pt}
\vskip 5pt
As an example, state~42 of Table~\ref{tbextend} looks like that:
\vskip 5pt
\ligne{\hfill
\small
$\vcenter{\vtop{\leftskip 0pt\parindent 0pt \hsize= 300pt
\ligne{\hfill\tt
XY1hhhhhZ010101Y01hhhhL010101Y11hhhhL010101Y001hhhR110101  
\hfill}
\ligne{\hfill\tt
 Y101hhhL010101Y011hhhZ010101Y111hhhL010101Y0001hhL010101  
\hfill}
\ligne{\hfill\tt
 Y1001hhZ010101Y0101hhZ010101Y1101hhL010101Y0011hhL010101  
\hfill}
\ligne{\hfill\tt
 Y1011hhL010101Y0111hhL010101Y1111hhZ010101Y00001hZ010101  
\hfill}
}}$
\hfill(8)\hskip 15pt}
\vskip 5pt
\ifnum 1 = 0 {
\noindent
which is perhaps better understandable if written like this:
\vskip 5pt
\ligne{\hfill\tt
1hhhhhZ010101  01hhhhL010101  11hhhhL010101  001hhhR110101  
\hfill}
\ligne{\hfill\tt
101hhhL010101  011hhhZ010101  111hhhL010101  0001hhL010101  
\hfill}
\ligne{\hfill\tt
1001hhZ010101  0101hhZ010101  1101hhL010101  0011hhL010101  
\hfill}
\ligne{\hfill\tt
1011hhL010101  0111hhL010101  1111hhZ010101  00001hZ010101  
\hfill}
\vskip 5pt
} \fi

The translation of~(8) into the RNA-alphabet is given in (9).
\vskip 5pt
\ligne{\hfill
\small
$\vcenter{\vtop{\leftskip 0pt\parindent 0pt \hsize= 300pt
\ligne{\hfill\tt
AACCCUGAGAGAGAGAAUCACUCACUCACUCCCACUGAGAGAGAACCACUCACUCACU
\hfill}
\ligne{\hfill\tt
  CCCUCUGAGAGAGAACCACUCACUCACUCCCACACUGAGAGAAGCUCUCACUCACU
\hfill}
\ligne{\hfill\tt
  CCCUCACUGAGAGAACCACUCACUCACUCCCACUCUGAGAGAAUCACUCACUCACU
\hfill}
\ligne{\hfill\tt
  CCCUCUCUGAGAGAACCACUCACUCACUCCCACACACUGAGAACCACUCACUCACU
\hfill}
\ligne{\hfill\tt
  CCCUCACACUGAGAAUCACUCACUCACUCCCACUCACUGAGAAUCACUCACUCACU
\hfill}
\ligne{\hfill\tt
  CCCUCUCACUGAGAACCACUCACUCACUCCCACACUCUGAGAACCACUCACUCACU
\hfill}
\ligne{\hfill\tt
  CCCUCACUCUGAGAACCACUCACUCACUCCCACUCUCUGAGAACCACUCACUCACU
\hfill}
\ligne{\hfill\tt
  CCCUCUCUCUGAGAAUCACUCACUCACUCCCACACACACUGAAUCACUCACUCACU
\hfill}
}}$
\hfill(9)\hskip 15pt}
\vskip 5pt
It is important to note that a code similar to (8) is dealt with by~$U$ but a code
similar to~(9) cannot be used by an RNA Turing machine. Let us see how such a universal 
RNA Turing machine works, say \RNAmtu, denote it by $rnaU$. It is assumed that the data 
used by $rnaU$ is constituted in the same way as (9), starting from (7). 
Table~\ref{tbRNA_mtu}, in Appendix 2, displays the program of $rnaU$. In the table,
as in Table~\ref{appPMTU}, the halting instructions are represented by an empty entry.
Moreover, as the program of Table~\ref{appPMTU}, the program has been checked by a 
computer program I devised to simulate it, as already mentioned.

The principle of transforming $U$ into $rnaU$ is simple: each state of~$U$ contains 16 
instructions. With a four-letter alphabet five states at most are needed to sort the 
instructions corresponding to those of~$U$. Indeed, the execution of an instruction
requires to write down two letters in place of the two ones scanned by the machine. That 
may require one, two, exceptionally four more states as a backward step may be required 
to overwrite a two symbol pattern. Most often a state is called by just a single other 
one from a single instruction. The calling instruction~$I$ has a move, 
say $\mu\in\{\LL,\RR\}$. Most often too, the majority of the instructions under the 
state~$s$ called by~$I$ have the same move~$\mu$. Most often too, in the calling state,
the calling instructions also have the same move. Consequently we have the following two 
patterns for those sets of instructions:
\vskip 5pt
\newdimen\larga\larga=14pt
\newdimen\largb\largb=21pt
\newdimen\largc\largc=140pt
\def\ivinst #1 #2 #3 #4 #5{
\ligne{\hfill
\hbox to \larga{\small\tt\hfill#1\hfill}\ \ 
\hbox to \largb{\small\tt\hfill#2\hfill} -\ 
\hbox to \largb{\small\tt\hfill#3\hfill} -\ 
\hbox to \largb{\small\tt\hfill#4\hfill} -\
\hbox to \largb{\small\tt\hfill#5\hfill} -\
\hfill}
}
\ligne{\hfill
\vtop{\leftskip 0pt\parindent 0pt\hsize=\largc
\ivinst {} A  C  G  U  
\vskip 1pt
\ivinst {001} {R002} {R003} {R004} {R005}
\vskip -2pt
\ivinst {002} {R00X} {R00L} {R00R} {R00Z} 
\vskip -2pt
\ivinst {003} {R00z} {R00Y} {R00\_} {R00l} 
\vskip -2pt
\ivinst {004} {R00h} {R00d} {R00U} {R00e}
\vskip -2pt
\ivinst {005} {R00T} {R00F} {R00W} {R00S}
}
\hfill
\vtop{\leftskip 0pt\parindent 0pt\hsize=\largc
\ivinst {} A  C  G  U  
\vskip 1pt
\ivinst {001} {L002} {L003} {L004} {L005}
\vskip -2pt
\ivinst {002} {L00X} {L00z} {L00h} {L00T}
\vskip -2pt
\ivinst {003} {L00L} {L00Y} {L00d} {L00F}
\vskip -2pt
\ivinst {004} {L00R} {L00\_} {L00U} {L00W} 
\vskip -2pt
\ivinst {005} {L00Z} {L00l} {L00e} {L00S}
}
\hfill\raise-33pt\hbox{(10)}\hskip 15pt}
\vskip 5pt
To left, in (10), we have the pattern for a move to right, to left of~(10), the patter 
for a move to left. Note that the 'instructions' are in fact pseudo-instructions which
should be replaced by the right state corresponding to the letter of~$U$ represented
by the position of the considered pseudo-instruction. In the case when several 
instructions with opposite moves call the same state, we need both sets together. In 
Table~\ref{tbRNA_mtu}, it happens once for states~{\tt 88} up to~{\tt 95} which simulate 
state~{\tt 21} of~$U$. But in several case, when the state only has two or three 
non-halting instructions, we need less states and also much less non halting 
instructions in the RNA version. In (11), (12) and (13) we give three examples
of the translation required for $rnaU$.

\ligne{\hfill
\vtop{\leftskip 0pt\parindent 0pt\hsize=310pt
\traitH
\vskip-2pt
\ligne{\hfill\lesvii {42} {\LL} {\LL} {{\tt R43}} {\LL} {\LL} {\LL} {\LL}
\addauxvii {\LL} {\LL} {\LL} \hfill}
\vskip-6pt
\traitH
}
\hfill}
\vskip 5pt
\ligne{\hfill
\vtop{\leftskip 0pt\parindent 0pt\hsize=\largc
\ivinst {} A  C  G  U  
\vskip 1pt
\ivinst {191} {L192} {L193} {L194} {L195}
\vskip -2pt
\ivinst {192} {}   {L191} {L191} {}
\vskip -2pt
\ivinst {193} {R194} {L191} {L191} {L191}
\vskip -2pt
\ivinst {194} {L191} {R196} {L191} {}
\vskip -2pt
\ivinst {195} {}   {L191} {L191} {}
}
\hfill\raise-33pt\hbox{(11)}\hskip 15pt}
\vskip 5pt
Those instructions belong to the Table~\ref{tbRNA_mtu}, but they are not the code of such 
a machine. We shall go back to that point in Sub-section~\ref{sbcontinue}. The piece of 
a table given in (11) implements the instructions of the state~{\tt 42} of~$U$ as 
described in Section~\ref{pedago}. Note that (11) applies to 16 instructions of~$U$ while 
Table~\ref{tbextend} given in Section~\ref{pedago} displays 10 instructions only for 
states~40 up to~45. That can be checked in Table~\ref{appPMTU} of Appendix~1 to the 
paper. In (12), we can see, to left, the transcription of state~{\tt 13}, to right, that 
of state~{\tt 14}. In Table~\ref{appPMTU}, both states have only two non halting 
instructions. In state~{\tt 13}, although the calling instruction and the instructions 
of state~{\tt 13} themselves
have \RR{} as motion, as \un{} is encoded \ftt{CU } and \dd{} is encoded \ftt{GC },
the rewriting of the scanned square requires a step to left in order to write down the 
appropriate letter followed by another step to right as the move is \RR. The same steps
are required in state~{\tt 14} to rewrite \un{} by \hh. But the second letter
in the RNA-code is the same for \un{} and \hh, which allows us to use the same RNA-state
for the step to right as there are halting instructions only for a first RNA-letter 
{\ftt A}. Hence two RNA-states are enough to translate state~{\tt 14} instead of three 
ones required to translate state~{\tt 13}.
\vskip 5pt
\ligne{\hfill
\vtop{\leftskip 0pt\parindent 0pt\hsize=\largc
\ivinst {} A  C  G  U  
\vskip 1pt
\ivinst {057} {} {R058}  {} {}
\vskip -2pt
\ivinst {058} {} {GR059} {} {CL058}
\vskip -2pt
\ivinst {059} {} {R060}  {} {}
}
\hfill
\vtop{\leftskip 0pt\parindent 0pt\hsize=\largc
\ivinst {} A  C  G  U  
\vskip 1pt
\ivinst {060} {}  {R061}  {} {}
\vskip -2pt
\ivinst {061} {R062} {GR060} {} {AL061}
}
\hfill\raise-20pt\hbox{(12)}\hskip 15pt}
\vskip 5pt
In (13), with have the unique case where the translation of a state of~$U$ requires
9 RNA-states. It concerns state~21 which is called by state~{\tt 17} and by 
state~{\tt 23}. In state~{\tt 17} the calling instruction is to left while it is to 
right under state~{\tt 23}.

\vskip 5pt
\ligne{\hfill
\vtop{\leftskip 0pt\parindent 0pt\hsize=\largc
\ivinst {} A  C  G  U  
\vskip 1pt
\ivinst {088} {G089}  {G090}  {G091}  {G092}
\ivinst {089} {}   {D102} {CG088} {CG090}
\ivinst {090} {GD097} {}   {G088}  {D102}
\ivinst {091} {GD091} {}   {AG095} {D089}
\ivinst {092} {}   {}   {D089}  {}
}
\hfill
\vtop{\leftskip 0pt\parindent 0pt\hsize=\largc
\ivinst {} A  C  G  U  
\vskip 1pt
\ivinst {093} {D094} {}   {D095}  {D096}
\ivinst {094} {}  {}   {AG090} {}
\ivinst {095} {G089} {G090}  {G097}  {CG090}
\ivinst {096} {}  {D102} {}   {}
}
\hfill\raise-30pt\hbox{(13)}\hskip 15pt}
\vskip 5pt
It can be seen that the instruction from state~{\tt 17} arrives on~\SS, while that from
state~{\tt 23} arrives on~\FF. 
\vskip 5pt
\ifnum 1 = 0 {
\ligne{\hfill\tt 
A\www C\www G\www U\www 
X\www Y\www 
L\www R\www Z\www V\www W\www 
F\www T\www h\www d\www e\hfill}
\ligne{\hfill\tt 
AA CC GG UU 
CA CU
AC AG AU UA UC 
UG CG GA GC GU\hfill}
} \fi
\vskip 5pt

   The number of letters of the code of~$U$ is 10,351 so that its translation in the
RNA-alphabet yields 20,702 letters. Note that the genome of the SARS-Cov2 virus contains 
around 31,000 letters, each one being materialised by a nucleotides. The genome of that
virus has 15 genes, we shall go back to that point in Sub-section~\ref{sbcontinue}
 
    If we consider the $RNA$-universal machine, its application to the $RNA$-code of the 
machine whose program is given in Table~\ref{tbaddcourt} requires 2,303,033 steps 
corresponding to the 106 steps of the courteous addition of that table. It was
checked by a computer program $\mathcal P$$_2$ I devised for that purpose.
The initial configuration of $rnaU$ is:
\vskip 5pt
\ligne{\hfill\footnotesize\tt
\vtop{\leftskip0pt\parindent 0pt\hsize=200pt
\ligne{\hfill UUUGCACUGAGAGGCUCUGAGAGGCUCUGAGAGGCUCUGAGA\hfill}
\ligne{\hfill GGCACUGAGAGGCUCUGAGAGGCUCUGAGAGGCACUGAGACG\hfill}
}\hfill}
\vskip 5pt
\noindent
and its final configuration is:
\vskip 5pt
\ligne{\hskip 84pt\footnotesize\tt
\vtop{\leftskip0pt\parindent 0pt\hsize=200pt
\ligne{UUUGCACUGAGAGGCUCUGAGAGGCUCUGAGAGGCUCUGAGA\hfill}
\ligne{GGCACUGAGAGGCUCUGAGAGGCUCUGAGAGGCACUGAGAGG\hfill}
\ligne{CUCUGAGAGGCUCUGAGAGGCUCUGAGAGGCUCUGAGA\hfill}
\ligne{GGCUCUGAGAGGCACUGAGACG\hfill}
}
\hfill}
\vskip 5pt
\noindent
In that final configuration, we can note that the head, {\ftt {UG }}, is on the square 
containing the
leftmost $*$ encoded as {\ftt {CACUGAGA }}. We can also check that there are five squares
{\ftt {GGCUCUGAGA }}, where {\ftt {CUCU }} encodes {\ftt{$|$ }}.

Compared to 
the 1,143,717 steps of the \PMTU, the $RNA$-universal machine requires a little more than
twice that number of steps.

\def\MM{\hbox{\tt M}}
   At this point, I have to indicate that it is easy to get a smaller universal Turing 
machine which can still be considered as pedagogic. There are two possible solutions which
can also be both applied. The first solution is to note that the construction of the
area to the left of~\FF{} with the unary pattern for the powers of~2 not greater 
than~$P$ requires 29 states. The price to pay is to require that the code of the simulated
Turing machine~$M$ contains the area constructed by~$U$ between the leftmost~\SS{} and
the rightmost \BB{} to the left-hand side of that \SS. The other solution consists in 
noticing that the majority of instructions in~$U$ are glides. If we append a new letter, 
say \MM, we can replace all halting instructions by the code {\small\tt MDM} where 
{\small\tt D} is the move and {\small\tt M} says that the element at the place of the 
symbol is the same as in the scanned square for the letter and the new state is the 
current one. 

    We conclude that subsection by indicating the RNA-translation of the small universal
Turing machine whose instructions are displayed by Table~\ref{tbneary}. We use (7) and
(14) displays the RNA-code which requires 410 letters.
\vskip 5pt
\ligne{\hfill\hskip 25pt
\vtop{\leftskip 0pt\parindent 0pt\hsize=180pt
\small\tt
\ligne{AACCCACACUGAACCUCCCUCUGAGAACCU\hfill}            
\ligne{CCCACUGAGAACCUCCCUGAGAGAAGCACU\hfill}            
\ligne{AACCCUCUGAGAAGCUCACUCCCUCUGAGAAGCACU\hfill}      
\ligne{CCCUCUGAGAAGCUCCCUGAGAGAAGCACU\hfill}            
\ligne{AACCCACACUGAACCUCUCCCACUGAGAAGCUCACU\hfill}      
\ligne{CCCACUGAGAACCUCUCCCACACUGAACCUCACU\hfill}        
\ligne{AACCCUGAGAGAAGCUCACUCCCUCUGAGAAGCACACU\hfill}    
\ligne{CCCACUGAGAAGCACUCCCAGAGAGAAGCACACU\hfill}        
\ligne{AACCCACUGAGAACCUCUCCCUCUGAGAAGCACUCU\hfill}      
\ligne{CCCACUGAGAACCACUCUCCCACACUGAAGCUCACU\hfill}      
\ligne{AACCCUGAGAGAAUCACUCUCCCUCUGAGAAGCUCACU\hfill}    
\ligne{CCCACUGAGAACCACACUCCCUCUGAGAAGCU\hfill}          
}
\hfill\raise-60pt\hbox{(14)}\hskip 15pt}
\vskip 5pt
   It is time to turn back to the dispute.

\ifnum 1=0 {
         A                  C                   G                  U

1:   001hL1              11hhL1                 01hhL1            1hhhR01
AACCCACACUGAACCU      CCCUCUGAGAACCU        CCCACUGAGAACCU    CCCUGAGAGAAGCACU
2:   11hhR101            11hhR01                11hhR1            1hhhR01
AACCCUCUGAGAAGCUCACU  CCCUCUGAGAAGCACU      CCCUCUGAGAAGCU    CCCUGAGAGAAGCACU
3:   001hL11             01hhR101              01hhL11            001hL101
AACCCACACUGAACCUCU   CCCACUGAGAAGCUCACU     CCCACUGAGAACCUCU  CCCACACUGAACCUCACU
4:   1hhhR101            11hhR001              01hhR01            1hhhR001
AACCCUGAGAGAAGCUCACU   CCCUCUGAGAAGCACACU   CCCACUGAGAAGCACU  CCCAGAGAGAAGCACACU
5:   01hhL11             11hhR011              01hhL011           001hR101
AACCCACUGAGAACCUCU   CCCUCUGAGAAGCACUCU     CCCACUGAGAACCACUCU  CCCACACUGAAGCUCACU
6:   1hhhZ011            11hhR101              01hhL001           11hhR1
AACCCUGAGAGAAUCACUCU   CCCUCUGAGAAGCUCACU   CCCACUGAGAACCACACU  CCCUCUGAGAAGCU
} \fi

\subsection{Continuation of the dispute}\label{sbcontinue}
 
    Clearly, the complexity of the RNA-code obtained from the code of~$U$ is significantly less than the size of the SARS-Cov2. It is comparable to the size of an influenza virus.
At first glance, as the behaviour of $U$ is unpredictable we could conclude that,
{\it a fortiori}, the behaviour of the SARS-Cov2 is also unpredictable. Although the
history of the pandemia confirms that point up to now, is that comparison of code sizes
a relevant argument?

    One can certainly raise the following objection: you speak of completely 
incomparable objects, what a Turing machine has to do with a virus?

    At first glance, the objection seems to be sound. Moreover, the tape of a Turing
machine is a linear thread like which is also the case of any code written within a
segment of the tape even if the segment does not contain any empty square. Indeed, a virus
has a complex $3D$-structure which is not at all represented by the Turing tape. It
can be objected to the objection that a Turing tape is an abstract object. What is 
important, in the tape, is that it has a linear order. Whether the tape is folded in
whichever way is meaningless for the computation. For RNA or DNA strands, the way they are
folded may be important as far as elements which are far away from each other in the
linear order may become neighbours thanks to the folding. That means more complication
on the side of viruses. Consequently, that objection does not seem to me a serious one.
Moreover, the $3D$-folding which makes molecules close neighbours has a kind of 
counter-parts in the Turing technology: in the \PMTU{} most instructions of a state
refer to a rather neighbouring states according to their numbering, if not the immediate
successor or the immediate predecessor. Other references to distant states which are 
less frequent may be looked as a jump which makes close states which are not according to
their numbering. Accordingly, the jumps perform what the folding does.

   Another objection could be the fact that in the concept of a Turing machine, at least
those considered in the present paper, the tape is infinite in both directions which has
no meaning in a biological context. I answered to that objection in underlying that
the configuration of a Turing machine is finite at each step of its computation. Remember 
that namely that property allows us to construct universal Turing machines. I already
stressed that point in the section devoted to the \PMTU. Now, a finite
segment which can be continued step by step as long as needed has its relevance in a
biological context. Another point on the side of the relevance is the consideration
of polite machines only as simulated machines. We mentioned that theoretically that
restriction does not alter the generality of the result as far as a non-polite machine
can be simulated by a polite one. But such a limitation on the tape means that it has
a beginning which is also the case for $RNA$-strands of a virus.

    The complexity is not the single argument in favour of the unpredictability of
the behaviour or evolution of viruses. The other reason I have to consider that
comparison as relevant lies in the behaviour of the \PMTU. What are its actions if
we only look at the behaviour of the head? We can mainly see two basic features.
In many cases, the head runs over the configuration looking after some pattern which is
expected to be somewhere on the tape. It is most probably something which also happens
within a cell during a replication process. That latter word points at the other main
activity of the Turing machine: it copies some portion of the tape over another portion
of the tape. What does a virus do? It takes advantage of the copying possibilities
contained in a cell to replicate itself as it cannot do that by itself only.
Copying and searching something are the basic actions of a Turing machine and those
phenomena are also at work at a bio molecular level. But there is a third point.
A code by itself can do nothing. It must be treated by a Turing machine in the case of
abstract computations, it must be treated by the cell machinery in case of the replication
of a virus. The computation of a Turing machine is performed in the head of the human 
being which created it. The same person may create a computer program in order to perform
that computation, in particular to check the correctness of the machine he/she built.
It is the reason why at several places I clearly distinguished between
a Turing machine and its encoding. As an example, (14) is a code which can be 
performed by the machine of which (11), (12) and (13) display a few states.
  
   There are simple model of computations, more simple but more abstract than the Turing
machine, which are based on that copying process like Post systems, for example,
see \cite{minsky} for a short description. I have 
no room to consider them here but it is significant that such models are used in order 
to make the machine of Table~\ref{tbneary} able of universal computations. Universality 
lies in the data too. Which makes us return to the code.

   Another point about Turing machines and viruses are the contrast with errors. When a
Turing machine works on some data, it is assumed that not only the data can be read by the
machine but also that there are no errors in the data. It is also assumed that there is
no error in the program of the Turing machine. In the abstract world in which they are 
living, Turing machines are error free objects which work on error free data without
error during the computation. A Turing machine exactly does what is written in its
program. Clearly, such a perfect behaviour is far away from what happens in a biological
context. However, things are not that different if we look at Turing machines as objects
created by human brains. How does a human being who decides to create a Turing machine
in order to solve a given problem? He/she writes down the program and submit it with
the data to a simulator of Turing machines. Whether the simulator is borrowed from
the web or it is created by the same person, there are several trials before the
human being obtains a machine he/she can trust as error free. Those trials are interesting
as far as there are usually errors during that tuning process. There are errors in the 
data which are not completely conformal to the format which they are supposed to obey.
There are also errors in the Turing machine which does not behave as expected. Sometimes,
the Turing machine halts because no instruction corresponds to what it reads under its
current state. Sometimes too, the Turing machine runs a so long time without halting
that its creator thinks it never halts. Looking carefully, the human being detects
an error in his/her program or in the data he/she wrote. Both cases may also happen 
together. Accordingly, data and program mute until they arrive to a sufficiently stable 
form. That real life process of creating a Turing machine looks like some evolution 
process of real life. Things occur as just described as far as nobody can prove the 
correctness of a program as far as the controlling tools are guaranteed to work just by 
long enough trials. We reach here the same limit to the power of universal Turing 
machine: there are (infinitely many) problems which they cannot solve. That latter 
sentence is a theorem.

   Another interesting point consists in the following statistics: among the 1652 
instructions of the \RNAmtu{} (4 letters and 413 states), 958 of them are glides: the 
head does 
not overwrite the scanned square. Also, among those 1652 instructions, 542 of them are
halting instructions, so that we remain with 152 instructions which overwrite the scanned
square with another letter: it means less than 10\%{} of the instructions make changes
on the tape. It looks like the situation in the genome where active coding elements are 
a small minority, which does not mean that the others are useless as is the case in 
the $RNA$ universal Turing machine: only 45 instructions among the non-halting one does
not change the state of the head. Moreover among those 45 instructions, 24 of them
replace the letter of the scanned square by another one. The big number of changes of 
states comes from the fact that the \PMTU~$U$ has 16 letters in its alphabet while
the \RNAmtu{} has only 4 of them at its disposal. In 
Table~\ref{tbRNA_mtu}, we dispatch the table of that latter machine by grouping its states
according to those of~$U$ which they are simulating. In each such groups, the
first state dispatches the execution into at most four states in order to sort what is
read according to the letter it represents in terms of those of~$U$. Note that the
operations of~$U$ can be split into nine parts, so that in mean, from 30 up to 53 states 
of the $RNA$-universal Turing machine correspond to a specific action. More precisely, 
those groups are the following:
\vskip 5pt
\ligne{\hfill
\footnotesize
1-53  ::  54-101  ::  102-132  ::  133-182  ::  183-231  :: 
232-281  ::  282-331  ::  332-373  ::  374-413
\hfill}
\vskip 5pt
The number of genes in a 
virus is very small. Let us remember that, as an example, the SARS-Cov2 has 15 of them.

   Assuming that the comparison I raised with the present paper is relevant, 
what can we infer from that?
   
   The first conclusion is the unpredictability of the evolution of a virus. All reasons
which were raised against the relevance of the comparison point at the same conclusion:
a virus is much more complex than a Turing machine so that if the behaviour of the latter
is unpredictable, it is all the more the case for the former. Does it mean that there is 
nothing to do to limit as strongly as possible the damage a virus can do when it is 
virulent? Certainly not. The comparison I consider means that something which looks like 
algorithmic has little chance to fight a virus. At that point, let us go back to the 
comparison. Computer scientists introduced sometimes a topology on data from which it 
can be proved that Turing machines viewed as operators
on the data are continuous. The notion of continuity corresponding to that topology can
be stated in a more concrete way: for a given Turing machine, if it is given long enough
data and if there are no differences between the data within that length,
the Turing machine gives the same output for those data. The larger that length with no
difference is, the finer, we say, the continuity is. In other words, if different data
entail the same behaviour of the machine, it means that the considered data are big and 
that 'close to the machine' we could say, they are not different. Every thing relies
on the fineness of that continuity.

    The consequences of that continuity theorem are that viruses try to cheat on
the immune system of their hosts, it is the way they can survive. When I say that 'the
virus tries' it is a short-circuit for the sentence: among the 'child'-virions produced 
by the 'parent'-virions, those which are the more adapted to the present situation
survive and only them. From the side of the immune system, its efficiency is a kind
of measure of the fineness of its continuity . If it can be cheated only when data are 
very large and are little different, then it will be difficult for a virus to deceive it. 
Viruses and our immune systems have a long history of constant struggle one against the
other.

    Homo sapiens sapiens is very proud of the knowledge that could be accumulated thanks
to science. Let us remember that the basis of today sciences stems from late 16$^{\rm th}$
century, mathematics being a bit older. Certainly it is a little duration at the 
geological scale. Indeed, that means less than 500 years, around 3000 years for 
mathematics. Life on our planet exists from over than 3,000,000,000 years. For sure, 
life evolution requires much time 
to evolve from viruses and bacteria to plants, to animals, to man. Let us stress that 
human beings are newcomers at geological scale. The way life appeared make all its 
components interact, and that was apparently always the case. Science of which homo 
sapiens sapiens is so proud is the work of how many people? Certainly less than one 
million and a half persons at world scale at the present moment and we have to compare 
that with the more than 7,000,000,000 human beings at present. Moreover, the number of 
to day scientists is several times more than the number of dead scientists since the 
beginning of science, at some point 5,000 years ago, roughly speaking. That large 
estimate shows us that scientific activity, whatever the appraisal of its results, is a 
very partial and very recent activity of mankind. Can it allow us to play with our 
environment as it were something given at our free use and free will? It seems to me 
that the answer is no. We should be careful when dealing with our environment. We should
try to foresee as far as possible the consequences of our actions on the environment.

Let us be more modest and let us try to take benefit of what can we learn from the other 
fields of science in each field of science. Theoretical computer scientists observe 
their environment and that careful look brought them valuable results. Perhaps Turing 
machine may be of help to better understand phenomena studied by other sciences. When 
computation and information are involved we may have relevant points of view for other 
people. Theoretical computer science bring some light to philosophical problems too, 
but that point goes for beyond the goal of that paper.

    I hope that the reader will find those comments of interest, whatever his/her point
of view.

\ifnum 1 = 0 {
1 h h h h h Z 0 1 0 1 0 1   0 1 h h h h L 0 1 0 1 0 1  
CUGAGAGAGAGAAUCACUCACUCACU  CACUGAGAGAGAACCACUCACUCACU
1 1 h h h h L 0 1 0 1 0 1   0 0 1 h h h R 1 1 0 1 0 1  
CUCUGAGAGAGAACCACUCACUCACU  CACACUGAGAGAAGCUCUCACUCACU
1 0 1 h h h L 0 1 0 1 0 1   0 1 1 h h h Z 0 1 0 1 0 1  
CUCACUGAGAGAACCACUCACUCACU  CACUCUGAGAGAAUCACUCACUCACU

1 1 1 h h h L 0 1 0 1 0 1   0 0 0 1 h h L 0 1 0 1 0 1
CUCUCUGAGAGAACCACUCACUCACU  CACACACUGAGAACCACUCACUCACU

1 0 0 1 h h Z 0 1 0 1 0 1   0 1 0 1 h h Z 0 1 0 1 0 1  
CUCACACUGAGAAUCACUCACUCACU  CACUCACUGAGAAUCACUCACUCACU

1 1 0 1 h h L 0 1 0 1 0 1   0 0 1 1 h h L 0 1 0 1 0 1  
CUCUCACUGAGAACCACUCACUCACU  CACACUCUGAGAACCACUCACUCACU

1 0 1 1 h h L 0 1 0 1 0 1   0 1 1 1 h h L 0 1 0 1 0 1  
CUCACUCUGAGAACCACUCACUCACU  CACUCUCUGAGAACCACUCACUCACU

1 1 1 1 h h Z 0 1 0 1 0 1   0 0 0 0 1 h Z 0 1 0 1 0 1
CUCUCUCUGAGAAUCACUCACUCACU  CACACACACUGAAUCACUCACUCACU

1hhhhhZ010101  01hhhhL010101  11hhhhL010101  001hhhR110101  101hhhL010101  011hhhZ010101  
111hhhL010101  0001hhL010101  1001hhZ010101  0101hhZ010101  1101hhL010101  0011hhL010101  
1011hhL010101  0111hhL010101  1111hhZ010101  00001hZ010101  
} \fi
\section{Conclusion}\label{conclude}

\newpage
\ligne{\hfill \bf Appendix 1\hfill}

\def\lesetatsv #1 #2 #3 #4 #5 #6{
\hfill\hbox to 18pt{\footnotesize\tt\hfill #1\hfill}\traitV
\hfill\hbox to 26pt{\footnotesize\tt\hfill #2\hfill}\traitV
\hfill\hbox to 26pt{\footnotesize\tt\hfill #3\hfill}\traitV
\hfill\hbox to 26pt{\footnotesize\tt\hfill #4\hfill}\traitV
\hfill\hbox to 26pt{\footnotesize\tt\hfill #5\hfill}\traitV
\hfill\hbox to 26pt{\footnotesize\tt\hfill #6\hfill}\traitV
}
\def\addlesvi #1 #2 #3 #4 #5 #6{
\hfill\hbox to 26pt{\footnotesize\tt\hfill #1\hfill}\traitV
\hfill\hbox to 26pt{\footnotesize\tt\hfill #2\hfill}\traitV
\hfill\hbox to 26pt{\footnotesize\tt\hfill #3\hfill}\traitV
\hfill\hbox to 26pt{\footnotesize\tt\hfill #4\hfill}\traitV
\hfill\hbox to 26pt{\footnotesize\tt\hfill #5\hfill}\traitV
\hfill\hbox to 26pt{\footnotesize\tt\hfill #6\hfill}\traitV
}

\ligne{\hfill
\vtop{\leftskip 0pt\parindent 0pt\hsize = 330pt
\begin{tab}\label{appPMTU}
\leurre
Table of $U$: it is devided in several sub-tables where the columns are labelled by 
the states and the lines are labelled by the letters. That latter display is more 
convenient, taking into account the number of states.
\vskip 3pt
\noindent
{\bf Sub-table~\ref{appPMTU}.a}: building the area between \SS{} and \FF. \TT{}
is missing as not used at that stage.
\end{tab}
\ligne{\hfill
\lesetatsv {} 1 2 3 4 5 \addlesvi 6 7 8 9 {10} {11} \hfill}
\vskip -6pt
\traitH
\vskip -2pt
\ligne{\hfill
\lesetatsv {\BB} {} {} {} {{\tt 0D05}} {} \addlesvi {} {{\tt 1R08}} {} {} {} 
{{\tt FR12}} \hfill}
\vskip -6pt
\traitH
\vskip -2pt
\ligne{\hfill
\lesetatsv {\zz} {\LL} {} {\RR} {\LL} {\RR} \addlesvi {\LL} {{\tt 1R08}} {\RR} {\RR} 
{\LL} {\LL} \hfill}
\vskip -6pt
\traitH
\vskip -2pt
\ligne{\hfill
\lesetatsv {\un} {\LL} {} {\RR} {\LL} {\RR} \addlesvi {\LL} {\LL} {\RR} {\RR} 
{\LL} {\LL} \hfill}
\vskip -6pt
\traitH
\vskip -2pt
\ligne{\hfill
\lesetatsv {\LL} {\LL} {} {\RR} {\LL} {\RR} \addlesvi {\LL} {} {\RR} {\RR} 
{\LL} {\LL} \hfill}
\vskip -6pt
\traitH
\vskip -2pt
\ligne{\hfill  
\lesetatsv {\RR} {\LL} {} {\RR} {\LL} {\RR} \addlesvi {\LL} {} {\RR} {\RR} 
{\LL} {\LL} \hfill}
\vskip -6pt
\traitH
\vskip -2pt
\ligne{\hfill
\lesetatsv {\XX} {\LL} {{\tt FR03}} {{\tt L06}} {} {\RR} \addlesvi {} {} {\RR} 
{{\tt FL10}}
{\LL} {\LL} \hfill}
\vskip -6pt
\traitH
\vskip -2pt
\ligne{\hfill
\lesetatsv {\YY} {\LL} {} {{\tt WL04}} {\LL} {\RR} \addlesvi {\LL} {} {\RR} {\RR} 
{\LL} {\LL} \hfill}
\vskip -6pt
\traitH
\vskip -2pt
\ligne{\hfill
\lesetatsv {\UU} {} {} {} {} {} \addlesvi {} {} {} {} 
{} {} \hfill}
\vskip -6pt
\traitH
\vskip -2pt
\ligne{\hfill
\lesetatsv {\WW} {{\tt SL}} {} {} {\LL} {{\tt YR03}} \addlesvi {} {} {} {} 
{} {} \hfill}
\vskip -6pt
\traitH
\vskip -2pt
\ligne{\hfill
\lesetatsv {\SS} {{\tt R02}} {} {} {\LL} {\RR} \addlesvi {} {\LL} {\RR} {{\tt L11}} 
{\tt L07} {\LL} \hfill}
\vskip -6pt
\traitH
\vskip -2pt
\ligne{\hfill
\lesetatsv {\hh} {\LL} {} {\RR} {\LL} {\RR} \addlesvi {\LL} {} {\RR} {\RR} 
{\LL} {\LL} \hfill}
\vskip -6pt
\traitH
\vskip -2pt
\ligne{\hfill
\lesetatsv {\dd} {} {} {} {} {} \addlesvi {} {} {} {} 
{} {} \hfill}
\vskip -6pt
\traitH
\vskip -2pt
\ligne{\hfill
\lesetatsv {\ee} {} {} {} {} {} \addlesvi {} {} {} {} 
{} {} \hfill}
\vskip -6pt
\traitH
\vskip -2pt
\ligne{\hfill
\lesetatsv {\FF} {} {} {\RR} {\LL} {\RR} \addlesvi {{\tt L07}} {} {{\tt XR09}} {} 
{} {} \hfill}
\vskip -6pt
\traitH
\vskip -2pt
\ligne{\hfill
\lesetatsv {\ZZ} {} {} {\RR} {\LL} {\RR} \addlesvi {\LL} {} {\RR} {\RR} 
{\LL} {\LL} \hfill}
\vskip -6pt
\traitH
}
\hfill}

\def\addlesv #1 #2 #3 #4 #5 {
\hfill\hbox to 26pt{\footnotesize\tt\hfill #1\hfill}\traitV
\hfill\hbox to 26pt{\footnotesize\tt\hfill #2\hfill}\traitV
\hfill\hbox to 26pt{\footnotesize\tt\hfill #3\hfill}\traitV
\hfill\hbox to 26pt{\footnotesize\tt\hfill #4\hfill}\traitV
\hfill\hbox to 26pt{\footnotesize\tt\hfill #5\hfill}\traitV
}
\vskip 15pt
\ligne{\hfill
\vtop{\leftskip 0pt\parindent 0pt\hsize = 330pt
{\leurre \it 
\noindent 
{\bf Sub-table~\ref{appPMTU}.b}:
States for constructing the scale of powers of two not greater than $P$.
The missing letters, namely \BB, \zz, \LL, \XX, \YY, \UU, \WW{} and \TT, involve 
halting instructions. It is the reason why the corresponding lines are missing in the
table.
}
\vskip 10pt
\ligne{\hfill
\lesetatsv {} {12} {13} {14} {15} {16} \addlesvi {17} {18} {19} {20} 
{21} {22}\hfill}
\vskip -6pt
\traitH
\vskip -2pt
\ligne{\hfill
\lesetatsv {\un} {d\RR 13} {d\RR 14} {h\RR 15} {d\LL 16} {} \addlesvi {h\LL 18} {} 
{\LL 20} {d\RR 16} {} {}\hfill}
\vskip -6pt
\traitH
\vskip -2pt
\ligne{\hfill
\lesetatsv {\RR} {} {} {} {} {\LL} \addlesvi {\RR} {\LL} {\RR} {} {h\LL 22} {h\LL}
\hfill}
\vskip -6pt
\traitH
\vskip -2pt
\ligne{\hfill
\lesetatsv {\SS} {} {} {} {} {} \addlesvi {\LL 21} {} {\LL 20} {} {} {}
\hfill}
\vskip -6pt
\traitH
\vskip -2pt
\ligne{\hfill
\lesetatsv {\hh} {} {} {} {} {R\RR 17} \addlesvi {\RR} {\LL} {\RR} {d\LL 16} {0\LL} {\LL} 
\hfill}
\vskip -6pt
\traitH
\vskip -2pt
\ligne{\hfill
\lesetatsv {\dd} {} {} {} {} {e\RR 17} \addlesvi {\RR} {e\RR 17} {} {} {\LL} {}
\hfill}
\vskip -6pt
\traitH
\vskip -2pt
\ligne{\hfill
\lesetatsv {\ee} {} {} {} {} {\LL} \addlesvi {\RR} {\LL 16} {\RR} {} {d\RR} {d\LL}
\hfill}
\vskip -6pt
\traitH
\vskip -2pt
\ligne{\hfill
\lesetatsv {\FF} {} {} {} {} {\RR 19} \addlesvi {} {} {} {} {\RR 23} {\RR 23}
\hfill}
\vskip -6pt
\traitH
}
\hfill}
\vskip 15pt
\def\lesetatsii #1 #2 #3{
\hfill\hbox to 12pt{\footnotesize\tt\hfill #1\hfill}\traitV
\hfill\hbox to 26pt{\footnotesize\tt\hfill #2\hfill}\traitV
\hfill\hbox to 26pt{\footnotesize\tt\hfill #3\hfill}\traitV
}
\noindent
{\leurre \it 
{\bf Sub-table~\ref{appPMTU}.c}:
To left: states for preparing the pattern to the left-hand side of \FF{} allowing
to convert the binary representation of~$n\leq P$ into $\hh^n$. The following symbols
are not concerned by the states~{\tt 23} up to~{\tt 27}: \LL, \RR, \XX, \YY, \UU, \WW,
\ee, \ZZ{} and \TT.
To right, on two sub-tables, the motion of the head to the \WW{} indicating the
scanned square on the tape of~$M$. Here too a few symbols are not concerned:
\BB, \dd, \ee, \FF{} and \TT.
}
\vskip 15pt
\ligne{\hfill
\vtop{\leftskip 0pt\parindent 0pt\hsize = 165pt
\ligne{\hfill
\lesetatsv {} {23} {24} {25} {26} {27} \hfill}
\vskip -6pt
\traitH
\vskip -2pt
\ligne{\hfill
\lesetatsv {\BB} {} {d\RR 25} {} {0\RR 25} {} \hfill}
\vskip -6pt
\traitH
\vskip -2pt
\ligne{\hfill
\lesetatsv {\zz} {1\RR} {\LL} {\RR} {\LL} {} \hfill}
\vskip -6pt
\traitH
\vskip -2pt
\ligne{\hfill
\lesetatsv {\un} {\RR} {\LL} {} {\LL} {0\LL} \hfill}
\vskip -6pt
\traitH
\vskip -2pt
\ligne{\hfill
\lesetatsv {\SS} {\LL 27} {} {} {} {} \hfill}
\vskip -6pt
\traitH
\vskip -2pt
\ligne{\hfill
\lesetatsv {\hh} {1\LL 26} {} {\RR} {} {} \hfill}
\vskip -6pt
\traitH
\vskip -2pt
\ligne{\hfill
\lesetatsv {\dd} {1\LL 24} {\LL} {\RR} {\LL} {} \hfill}
\vskip -6pt
\traitH
\vskip -2pt
\ligne{\hfill
\lesetatsv {\FF} {\RR 21} {\LL} {\RR 23} {\LL} {\RR 28} \hfill}
\vskip -6pt
\traitH
}
\hfill
\vtop{\leftskip 0pt\parindent 0pt\hsize = 84pt
\ligne{\hfill
\lesetatsii {} {28} {29} \hfill}
\vskip -6pt
\traitH
\vskip-2pt
\ligne{\hfill
\lesetatsii {\zz} {\RR} {\RR} \hfill}
\vskip -6pt
\traitH
\vskip-2pt
\ligne{\hfill
\lesetatsii {\un} {\RR} {\RR} \hfill}
\vskip -6pt
\traitH
\vskip-2pt
\ligne{\hfill
\lesetatsii {\LL} {\RR} {\RR} \hfill}
\vskip -6pt
\traitH
\vskip-2pt
\ligne{\hfill
\lesetatsii {\RR} {\RR} {\RR} \hfill}
\vskip -6pt
\traitH
\vskip-2pt
\ligne{\hfill
\lesetatsii {\XX} {W\RR 29} {\RR} \hfill}
\vskip -6pt
\traitH
\vskip-2pt
\ligne{\hfill
\lesetatsii {\YY} {\RR} {\RR} \hfill}
\vskip -6pt
\traitH
}
\hskip 0pt
\vtop{\leftskip 0pt\parindent 0pt\hsize = 84pt
\ligne{\hfill
\lesetatsii {} {28} {29} \hfill}
\vskip -6pt
\traitH
\vskip-2pt
\ligne{\hfill
\lesetatsii {\UU} {}   {\RR} \hfill}
\vskip -6pt
\traitH
\vskip-2pt
\ligne{\hfill
\lesetatsii {\WW} {}   {\RR 30} \hfill}
\vskip -6pt
\traitH
\vskip-2pt
\ligne{\hfill
\lesetatsii {\SS} {\RR} {\RR} \hfill}
\vskip -6pt
\traitH
\vskip-2pt
\ligne{\hfill
\lesetatsii {\hh} {}   {\RR} \hfill}
\vskip -6pt
\traitH
\vskip-2pt
\ligne{\hfill
\lesetatsii {\ZZ} {\RR} {\RR} \hfill}
\vskip -6pt
\traitH
}
\hfill
}
\vskip 15pt
\ligne{\hfill
\vtop{\leftskip 0pt\parindent 0pt\hsize = 330pt
{\leurre \it 
{\bf Sub-table~\ref{appPMTU}.d}:
States for copying the binary representation of $n$ in the scanned square of the tape 
of~$M$ onto the pattern of the useful powers of two. Here there is no missing letter
as far as each letter is read under at least one state among those of that part of
the table.
}
\vskip 10pt
\ligne{\hfill
\lesetatsv {} {30} {31} {32} {33} {34} \addlesv {35} {36} {37} {38} {39} \hfill}
\vskip -6pt
\traitH
\vskip -2pt
\ligne{\hfill
\lesetatsv {\BB} {} {} {} {} {} \addlesv {} {} {} {\RR{} 39} {} \hfill}
\vskip -6pt
\traitH
\vskip -2pt
\ligne{\hfill
\lesetatsv {\zz} {d\LL 31} {\LL} {\LL} {\RR} {\RR} \addlesv {\LL} {\LL} {\LL} 
{\LL} {\RR} \hfill}
\vskip -6pt
\traitH
\vskip -2pt
\ligne{\hfill
\lesetatsv {\un} {e\LL 35} {\LL} {} {\RR} {\RR} \addlesv {\LL} {} {\LL} 
{\LL} {} \hfill}
\vskip -6pt
\traitH
\vskip -2pt
\ligne{\hfill
\lesetatsv {\LL} {} {\LL} {} {\RR} {\RR} \addlesv {\LL} {} {\LL} 
{} {} \hfill}
\vskip -6pt
\traitH
\vskip -2pt
\ligne{\hfill
\lesetatsv {\RR} {} {\LL} {} {\RR} {\RR} \addlesv {\LL} {} {\LL} 
{} {} \hfill}
\vskip -6pt
\traitH
\vskip -2pt
\ligne{\hfill
\lesetatsv {\XX} {\LL 37} {\LL} {} {\RR} {\RR} \addlesv {\LL} {} {\LL} {} {} \hfill}
\vskip -6pt
\traitH
\vskip -2pt
\ligne{\hfill
\lesetatsv {\YY} {\LL 37} {\LL} {} {\RR} {\RR} \addlesv {\LL} {} {\LL} {} {} \hfill}
\vskip -6pt
\traitH
\vskip -2pt
\ligne{\hfill
\lesetatsv {\UU} {} {\LL} {} {} {\RR} \addlesv {\LL} {} {\LL} {} {} \hfill}
\vskip -6pt
\traitH
\vskip -2pt
\ligne{\hfill
\lesetatsv {\WW} {} {\LL} {} {\RR{} 34} {\RR{} 30} \addlesv {\LL} {} {\LL} {} {} \hfill}
\vskip -6pt
\traitH
\vskip -2pt
\ligne{\hfill
\lesetatsv {\SS} {} {\LL} {} {\RR} {\RR} \addlesv {\LL} {} {\LL} {} {} \hfill}
\vskip -6pt
\traitH
\vskip -2pt
\ligne{\hfill
\lesetatsv {\hh} {\LL 37} {\LL} {\LL} {\RR} {\RR} \addlesv {\LL} {\LL} {\LL} 
{\LL} {\RR} \hfill}
\vskip -6pt
\traitH
\vskip -2pt
\ligne{\hfill
\lesetatsv {\dd} {\RR} {\LL} {h\RR 33} {} {0\RR 30} \addlesv {\LL} {e\RR 33} {0\LL} 
{\LL} {\RR} \hfill}
\vskip -6pt
\traitH
\vskip -2pt
\ligne{\hfill
\lesetatsv {\ee} {\RR} {\LL} {\LL} {\RR} {1\RR 30} \addlesv {\LL} {\LL} {1\LL} 
{\LL} {L\RR 40} \hfill}
\vskip -6pt
\traitH
\vskip -2pt
\ligne{\hfill
\lesetatsv {\FF} {} {\LL 32} {} {\RR} {} \addlesv {\LL 36} {} {\LL 38} {} {} \hfill}
\vskip -6pt
\traitH
\vskip -2pt
\ligne{\hfill
\lesetatsv {\ZZ} {} {\LL} {} {\RR} {\RR} \addlesv {\LL} {} {\LL} {} {} \hfill}
\vskip -6pt
\traitH
\vskip -2pt
\ligne{\hfill
\lesetatsv {\TT} {} {\LL} {} {\RR} {\RR} \addlesv {\LL} {} {\LL} {} {} \hfill}
\vskip -6pt
\traitH
}
\hfill}
\vskip 15pt
\ligne{\hfill
\vtop{\leftskip 0pt\parindent 0pt\hsize = 330pt
{\leurre \it 
{\bf Sub-table~\ref{appPMTU}.e}:
Left-hand side half, states~{\tt 40} up to~{\tt 45}: transformation of the binary 
representation of~$n$ copied to the right-hand side of \FF{} into $n$ copies of~\hh{}
to the right-hand side of \FF. Those states are used twice during the simulation of one
step of $M$ computation.
\vskip 1pt
Right-hand side half, states~{\tt 45} up to~{\tt 46}: thanks to the unary representation
of~$n$, location of the instruction to be applied under the state marked by \WW.
}
\vskip 5pt
\ligne{\hfill
\lesetatsv {} {40} {41} {42} {43} {44} \addlesvi {45} {46} {47} {48} {49} {50} \hfill}
\vskip -6pt
\traitH
\vskip-2pt
\ligne{\hfill
\lesetatsv {\zz} {\RR} {h\LL 42} {\LL} {Y\RR 40} {} \addlesvi {\RR} {\RR} {\RR} {\LL} 
 {\LL} {\RR} \hfill}
\vskip -6pt
\traitH
\vskip-2pt
\ligne{\hfill
\lesetatsv {\un} {} {\LL} {\LL} {} {} \addlesvi {} {\RR} {\RR} {\LL} {} {\RR} \hfill}
\vskip -6pt
\traitH
\vskip-2pt
\ligne{\hfill
\lesetatsv {\LL} {} {} {\RR 43} {} {d\RR 45} \addlesvi {} {\RR} {\RR} {\LL} {} {\RR} 
\hfill}
\vskip -6pt
\traitH
\vskip-2pt
\ligne{\hfill
\lesetatsv {\RR} {} {} {\LL} {\RR} {e\LL} \addlesvi {} {\RR} {\RR} {\LL} {} {\RR} \hfill}
\vskip -6pt
\traitH
\vskip-2pt
\ligne{\hfill
\lesetatsv {\XX} {} {} {} {} {} \addlesvi {} {\RR} {} {\LL} {} {\RR} \hfill}
\vskip -6pt
\traitH
\vskip-2pt
\ligne{\hfill
\lesetatsv {\YY} {} {} {\LL} {\RR} {0\LL} \addlesvi {} {\RR} {F\LL 48} {\LL} {} {\RR} 
\hfill}
\vskip -6pt
\traitH
\vskip-2pt
\ligne{\hfill
\lesetatsv {\UU} {\RR} {} {\LL} {\RR} {d\LL} \addlesvi {} {} {} {} {} {} \hfill}
\vskip -6pt
\traitH
\vskip-2pt
\ligne{\hfill
\lesetatsv {\WW} {} {} {} {} {} \addlesvi {} {\RR 47} {} {\LL} {} {\RR} \hfill}
\vskip -6pt
\traitH
\vskip-2pt
\ligne{\hfill
\lesetatsv {\SS} {} {} {} {} {} \addlesvi {} {\RR} {} {\LL 49} {} {\RR} \hfill}
\vskip -6pt
\traitH
\vskip-2pt
\ligne{\hfill
\lesetatsv {\hh} {\RR} {\RR} {\LL} {U\RR 40} {} \addlesvi {0\RR 46} {\RR} {\RR} {\LL} 
  {0\RR 50} {\RR} \hfill}
\vskip -6pt
\traitH
\vskip-2pt
\ligne{\hfill
\lesetatsv {\dd} {h\RR} {} {\LL} {} {} \addlesvi {\RR} {} {} {} {} {} \hfill}
\vskip -6pt
\traitH
\vskip-2pt
\ligne{\hfill
\lesetatsv {\ee} {\RR} {} {\LL} {R\RR 40} {} \addlesvi {L\RR 40} {} {} {} {} {} \hfill}
\vskip -6pt
\traitH
\vskip-2pt
\ligne{\hfill
\lesetatsv {\FF} {\RR 41} {\LL 42} {\LL} {\LL 44} {} \addlesvi {\RR} {} {} {} {\RR 51} 
 {Y\RR 47} \hfill}
\vskip -6pt
\traitH
\vskip-2pt
\ligne{\hfill
\lesetatsv {\ZZ} {} {} {} {\LL 84} {} \addlesvi {} {\RR} {\RR} {\LL} {} {\RR} \hfill}
\vskip -6pt
\traitH
\vskip-2pt
\ligne{\hfill
\lesetatsv {\TT} {} {} {} {} {} \addlesvi {} {\RR 80} {} {} {} {} \hfill}
\vskip -6pt
\traitH
}
\hfill}
\vskip 10pt
\ligne{\hfill
\vtop{\leftskip 0pt\parindent 0pt\hsize = 330pt
{\leurre \it 
{\bf Sub-table~\ref{appPMTU}.f}: States~{\tt 51} up to~{\tt 60}. Copying the letter
of the instruction onto the square scanned by~$M$.
}
\vskip 5pt
\ligne{\hfill
\lesetatsv {}  {51} {52} {53} {54} {55} \addlesv {56} {57} {58} {59} {60} \hfill}
\vskip -6pt
\traitH
\vskip-2pt
\ligne{\hfill
\lesetatsv {\zz} {\RR} {d\RR 53} {\RR} {d\LL 55} {} \addlesv {\LL} {\RR} {e\LL 59} 
 {} {\LL} \hfill}
\vskip -6pt
\traitH
\vskip-2pt
\ligne{\hfill
\lesetatsv {\un} {\RR} {e\RR 57} {\RR} {d\LL 55} {} \addlesv {\LL} {\RR} {e\LL 59} 
 {} {\LL} \hfill}
\vskip -6pt
\traitH
\vskip-2pt
\ligne{\hfill
\lesetatsv {\LL} {\RR} {} {\RR} {} {} \addlesv {\LL} {\RR} {} {} {\LL} \hfill}
\vskip -6pt
\traitH
\vskip-2pt
\ligne{\hfill
\lesetatsv {\RR} {\RR} {} {\RR} {} {} \addlesv {\LL} {\RR} {} {} {\LL} \hfill}
\vskip -6pt
\traitH
\vskip-2pt
\ligne{\hfill
\lesetatsv {\XX} {\RR} {} {\RR} {} {} \addlesv {\LL} {\RR} {} {} {\LL} \hfill}
\vskip -6pt
\traitH
\vskip-2pt
\ligne{\hfill
\lesetatsv {\YY} {\RR} {\RR} {\RR} {} {} \addlesv {\LL} {\RR} {} {} {\LL} \hfill}
\vskip -6pt
\traitH
\vskip-2pt
\ligne{\hfill
\lesetatsv {\UU} {} {} {\RR} {} {} \addlesv {\LL} {\RR} {} {} {\LL} \hfill}
\vskip -6pt
\traitH
\vskip-2pt
\ligne{\hfill
\lesetatsv {\WW} {\RR} {} {\RR 54} {} {\LL 56} \addlesv {\RR 52} {\RR 58} {} {\LL 60}
 {\RR 52} \hfill}
\vskip -6pt
\traitH
\vskip-2pt
\ligne{\hfill
\lesetatsv {\SS} {\RR} {} {\RR} {} {} \addlesv {\LL} {\RR} {} {\LL} {\LL} \hfill}
\vskip -6pt
\traitH
\vskip-2pt
\ligne{\hfill
\lesetatsv {\hh} {\RR} {\LL 61} {\RR} {d\LL 55} {} \addlesv {\LL} {\RR} {e\LL 59} 
 {\LL} {\LL} \hfill}
\vskip -6pt
\traitH
\vskip-2pt
\ligne{\hfill
\lesetatsv {\dd} {} {\RR} {} {\RR} {\LL} \addlesv {\RR 52} {} {\RR} {\LL} {\LL} \hfill}
\vskip -6pt
\traitH
\vskip-2pt
\ligne{\hfill
\lesetatsv {\ee} {} {\RR} {} {\RR} {\LL} \addlesv {\RR 52} {} {\RR} {\LL} {\LL} \hfill}
\vskip -6pt
\traitH
\vskip-2pt
\ligne{\hfill
\lesetatsv {\FF} {W\RR 52} {} {} {} {} \addlesv {} {} {} {} {\LL} \hfill}
\vskip -6pt
\traitH
\vskip-2pt
\ligne{\hfill
\lesetatsv {\ZZ} {\RR} {} {\RR} {} {} \addlesv {\LL} {\RR} {} {} {\LL} \hfill}
\vskip -6pt
\traitH
}
\hfill}
\vskip 10pt
\ligne{\hfill
\vtop{\leftskip 0pt\parindent 0pt\hsize = 330pt
{\leurre \it 
{\bf Sub-table~\ref{appPMTU}.g}: States~{\tt 61} up to~{\tt 70}. Return to the scanned
square in order to erase the markings and then return to clear the instruction from
the markings. Then, $U$ performs the move of the head of~$M$ over its tape.
}
\vskip 5pt
\ligne{\hfill
\lesetatsv {}  {61} {62} {63} {64} {65} \addlesv {66} {67} {68} {69} {70} \hfill}
\vskip -6pt
\traitH
\vskip-2pt
\ligne{\hfill
\lesetatsv {\BB} {} {} {} {} {} \addlesv {} {} {} {} {U\LL 71} \hfill}
\vskip -6pt
\traitH
\vskip-2pt
\ligne{\hfill
\lesetatsv {\zz} {\LL} {\RR} {h\RR} {\LL} {\LL} \addlesv {\RR} {\RR} {\LL} {\RR} {\RR} 
\hfill}
\vskip -6pt
\traitH
\vskip-2pt
\ligne{\hfill
\lesetatsv {\un} {\LL} {\RR} {h\RR} {\LL} {\LL} \addlesv {\RR} {\RR} {\LL} {\RR} {\RR} 
\hfill}
\vskip -6pt
\traitH
\vskip-2pt
\ligne{\hfill
\lesetatsv {\LL} {} {\RR} {} {\LL} {\LL} \addlesv {U\RR 67} {\RR} {} {\RR} {\RR} \hfill}
\vskip -6pt
\traitH
\vskip-2pt
\ligne{\hfill
\lesetatsv {\RR} {} {\RR} {} {\LL} {\LL} \addlesv {F\RR 69} {\RR} {} {\RR} {\RR} \hfill}
\vskip -6pt
\traitH
\vskip-2pt
\ligne{\hfill
\lesetatsv {\XX} {} {\RR} {} {\LL} {\LL} \addlesv {} {\RR} {} {\RR} {} \hfill}
\vskip -6pt
\traitH
\vskip-2pt
\ligne{\hfill
\lesetatsv {\YY} {} {\RR} {} {\LL} {\LL} \addlesv {} {\RR} {} {\RR} {} \hfill}
\vskip -6pt
\traitH
\vskip-2pt
\ligne{\hfill
\lesetatsv {\UU} {} {\RR} {} {\LL} {\LL} \addlesv {} {\RR} {W\LL 76} {\RR} {W\LL 76} 
\hfill}
\vskip -6pt
\traitH
\vskip-2pt
\ligne{\hfill
\lesetatsv {\WW} {\RR 62} {\RR 63} {} {\LL 65} {\RR 66} \addlesv {} {U\LL 68} {} 
 {U\RR 70} {} \hfill}
\vskip -6pt
\traitH
\vskip-2pt
\ligne{\hfill
\lesetatsv {\SS} {} {\RR} {} {\LL} {\LL} \addlesv {} {\RR} {} {\RR} {} \hfill}
\vskip -6pt
\traitH
\vskip-2pt
\ligne{\hfill
\lesetatsv {\hh} {} {\RR} {\LL 64} {\LL} {\LL} \addlesv {\RR} {\RR} {\LL} {\RR} {\RR} 
\hfill}
\vskip -6pt
\traitH
\vskip-2pt
\ligne{\hfill
\lesetatsv {\dd} {0\LL} {} {0\RR} {\LL} {\LL} \addlesv {} {} {} {} {} \hfill}
\vskip -6pt
\traitH
\vskip-2pt
\ligne{\hfill
\lesetatsv {\ee} {1\LL} {} {1\RR} {\LL} {\LL} \addlesv {} {} {} {} {} \hfill}
\vskip -6pt
\traitH
\vskip-2pt
\ligne{\hfill
\lesetatsv {\ZZ} {} {\RR} {} {\LL} {\LL} \addlesv {} {\RR} {} {\RR} {\RR} \hfill}
\vskip -6pt
\traitH
}
\hfill}
\vskip 15pt
\def\addlesiv #1 #2 #3 #4{
\hfill\hbox to 26pt{\footnotesize\tt\hfill #1\hfill}\traitV
\hfill\hbox to 26pt{\footnotesize\tt\hfill #2\hfill}\traitV
\hfill\hbox to 26pt{\footnotesize\tt\hfill #3\hfill}\traitV
\hfill\hbox to 26pt{\footnotesize\tt\hfill #4\hfill}\traitV
}
\ligne{\hfill
\vtop{\leftskip 0pt\parindent 0pt\hsize = 300pt
{\leurre \it 
{\bf Sub-table~\ref{appPMTU}.h}: States~{\tt 71} up to~{\tt 79}. Creating a new
empty square at the right-hand side end of the tape of~$M$. The first states also clear
the marking performed during the action indicated by Sub-table~{\bf \ref{appPMTU}.g}.
We also have a new call to states~{\tt 30} up to~{\tt 45} to copy the binary 
representation of the new state onto the scale-pattern to the left-hand side of~\FF{}
and to compute to the right-hand side of it the unary representation of that number.
}
\vskip 5pt
\ligne{\hfill
\lesetatsv {}  {71} {72} {73} {74} {75} \addlesiv {76} {77} {78} {79} \hfill}
\vskip -6pt
\traitH
\vskip-2pt
\ligne{\hfill
\lesetatsv {\BB} {} {h\LL 73} {} {} {} \addlesiv {} {} {} {} \hfill}
\vskip -6pt
\traitH
\vskip-2pt
\ligne{\hfill
\lesetatsv {\zz}  {} {} {} {e\RR 72} {} \addlesiv {\LL} {\LL} {\RR} {\RR} \hfill}
\vskip -6pt
\traitH
\vskip-2pt
\ligne{\hfill
\lesetatsv {\un}  {} {} {} {R\RR 72} {\RR} \addlesiv {\LL} {\LL} {\RR} {\RR} \hfill}
\vskip -6pt
\traitH
\vskip-2pt
\ligne{\hfill
\lesetatsv {\LL}  {} {} {} {} {} \addlesiv {} {\LL} {\RR} {\RR} \hfill}
\vskip -6pt
\traitH
\vskip-2pt
\ligne{\hfill
\lesetatsv {\RR}  {} {\RR} {} {\LL} {1\RR} \addlesiv {} {\LL} {\RR} {\RR} \hfill}
\vskip -6pt
\traitH
\vskip-2pt
\ligne{\hfill
\lesetatsv {\XX}  {} {} {} {} {} \addlesiv {} {\LL} {\RR} {\RR} \hfill}
\vskip -6pt
\traitH
\vskip-2pt
\ligne{\hfill
\lesetatsv {\YY}  {} {\RR} {} {\LL} {h\RR} \addlesiv {\LL} {\LL} {\RR} {\RR} \hfill}
\vskip -6pt
\traitH
\vskip-2pt
\ligne{\hfill
\lesetatsv {\UU}  {} {\RR} {\LL 74} {\RR 75} {W\RR} \addlesiv {\LL} {\LL} {} {L\RR 30} 
\hfill}
\vskip -6pt
\traitH
\vskip-2pt
\ligne{\hfill
\lesetatsv {\WW}  {} {} {} {} {} \addlesiv {\LL} {\LL} {X\RR 79} {\RR} \hfill}
\vskip -6pt
\traitH
\vskip-2pt
\ligne{\hfill
\lesetatsv {\SS}  {} {} {} {} {} \addlesiv {\LL 77} {T\RR 78} {} {} \hfill}
\vskip -6pt
\traitH
\vskip-2pt
\ligne{\hfill
\lesetatsv {\hh}  {Y\RR 72} {\RR} {\LL} {Y\RR 72} {1\LL 76} \addlesiv {\LL} {\LL} 
 {\RR} {\RR} \hfill}
\vskip -6pt
\traitH
\vskip-2pt
\ligne{\hfill
\lesetatsv {\ee}  {} {\RR} {} {\LL} {0\RR} \addlesiv {} {} {} {} \hfill}
\vskip -6pt
\traitH
\vskip-2pt
\ligne{\hfill
\lesetatsv {\FF}  {} {} {} {} {} \addlesiv {} {\LL} {} {R\RR 30} \hfill}
\vskip -6pt
\traitH
\vskip-2pt
\ligne{\hfill
\lesetatsv {\ZZ}  {} {} {} {} {} \addlesiv {} {\LL} {\RR} {\RR} \hfill}
\vskip -6pt
\traitH
}
\hfill}
\def\addlesiii #1 #2 #3{
\hfill\hbox to 26pt{\footnotesize\tt\hfill #1\hfill}\traitV
\hfill\hbox to 26pt{\footnotesize\tt\hfill #2\hfill}\traitV
\hfill\hbox to 26pt{\footnotesize\tt\hfill #3\hfill}\traitV
}
\vskip 15pt
\ligne{\hfill
\vtop{\leftskip 0pt\parindent 0pt\hsize = 300pt
{\leurre \it 
{\bf Sub-table~\ref{appPMTU}.i}: States~{\tt 70} up to~{\tt 87}. Locating the new state
and, when it is obtained, start the simulation of the next step of~$M$ in its
computation.
}
\vskip 5pt
\ligne{\hfill
\lesetatsv {}  {80} {81} {82} {83} {84} \addlesiii {85} {86} {87} \hfill}
\vskip -6pt
\traitH
\vskip -2pt
\ligne{\hfill
\lesetatsv {\BB}  {} {} {} {} {\RR 85} \addlesiii {} {} {} \hfill}
\vskip -6pt
\traitH
\vskip -2pt
\ligne{\hfill
\lesetatsv {\zz}  {\RR} {\RR} {\RR} {\RR} {\LL} \addlesiii {\RR} {\RR} {\LL} \hfill}
\vskip -6pt
\traitH
\vskip -2pt
\ligne{\hfill
\lesetatsv {\un}  {\RR} {} {\RR} {\RR} {\LL} \addlesiii {} {\RR} {\LL} \hfill}
\vskip -6pt
\traitH
\vskip -2pt
\ligne{\hfill
\lesetatsv {\LL}  {\RR} {} {\RR} {\RR} {\LL} \addlesiii {} {\RR} {\LL} \hfill}
\vskip -6pt
\traitH
\vskip -2pt
\ligne{\hfill
\lesetatsv {\RR}  {\RR} {} {\RR} {\RR} {\LL} \addlesiii {} {\RR} {\LL} \hfill}
\vskip -6pt
\traitH
\vskip -2pt
\ligne{\hfill
\lesetatsv {\XX}  {F\LL 84} {} {\RR} {F\LL 84} {\LL} \addlesiii {} {\RR} {\LL} \hfill}
\vskip -6pt
\traitH
\vskip -2pt
\ligne{\hfill
\lesetatsv {\YY}  {\RR} {} {\RR} {\RR} {\LL} \addlesiii {} {\RR} {\LL} \hfill}
\vskip -6pt
\traitH
\vskip -2pt
\ligne{\hfill
\lesetatsv {\WW}  {} {} {} {Y\RR} {\LL} \addlesiii {} {Y\RR} {\RR 29} \hfill}
\vskip -6pt
\traitH
\vskip -2pt
\ligne{\hfill
\lesetatsv {\SS}  {} {} {} {} {} \addlesiii {} {\LL 87} {} \hfill}
\vskip -6pt
\traitH
\vskip -2pt
\ligne{\hfill
\lesetatsv {\hh}  {\RR} {\RR} {\RR} {\RR} {\LL} \addlesiii {0\RR 81} {\RR} {\LL} \hfill}
\vskip -6pt
\traitH
\vskip -2pt
\ligne{\hfill
\lesetatsv {\dd}  {} {} {} {} {\LL} \addlesiii {\RR} {} {} \hfill}
\vskip -6pt
\traitH
\vskip -2pt
\ligne{\hfill
\lesetatsv {\FF}  {} {} {X\RR 83} {} {\LL} \addlesiii {\RR} {W\RR} {} \hfill}
\vskip -6pt
\traitH
\vskip -2pt
\ligne{\hfill
\lesetatsv {\ZZ}  {\RR} {} {\RR} {\RR} {\LL} \addlesiii {} {\RR} {\LL} \hfill}
\vskip -6pt
\traitH
\vskip -2pt
\ligne{\hfill
\lesetatsv {\TT}  {} {\RR 82} {} {} {\LL} \addlesiii {S\RR 86} {\RR} {} \hfill}
\vskip -6pt
\traitH
}
\hfill}
\newpage
\ligne{\hfill \bf Appendix 2\hfill}
\vskip 10pt

\begin{tab}\label{tbRNA_mtu}
\leurre
The $RNA$-universal Turing machine: it works with the encoding defined both by
that of the \PMTU{} and by its translation in the $RNA$-alphabet. The label
{\tt state $n$} indicates the state of the \PMTU{} which is simulated by the states of
the $RNA$-universal Turing machine which follows it. Empty entries are halting 
instructions.
\end{tab}

\newcount\stnumer\stnumer=1
\largb=24pt
\largc=145pt
\ligne{\hfill
\vtop{\leftskip 0pt\parindent 0pt\hsize=\largc\baselineskip 8pt
\ivinst {} A  C  G  U  
\vskip 2pt
\ligne{\hrulefill}
\vskip 2pt
\ligne{\tt state \the\stnumer \global\advance \stnumer by 1\hfill}
\vskip 2pt
\ivinst {  1} {L2} {L3} {L4} {L5} 
\ivinst {  2} {L1} {L1} {L1} {R6} 
\ivinst {  3} {L1} {L1} {UL} {L1} 
\ivinst {  4} {L1} {} {} {R3} 
\ivinst {  5} {} {L1} {} {R2} 
\vskip 2pt
\ligne{\tt state \the\stnumer \global\advance \stnumer by 1\hfill}
\vskip 2pt
\ivinst {  6} {R7} {} {} {} 
\ivinst {  7} {CL8} {R9} {} {} 
\ivinst {  8} {UR7} {} {} {} 
\vskip 2pt
\ligne{\tt state \the\stnumer \global\advance \stnumer by 1\hfill}
\vskip 2pt
\ivinst {  9} {R10} {R11} {R12} {R13} 
\ivinst { 10} {L13} {R9} {R9} {R9} 
\ivinst { 11} {R9} {GL12} {} {R9} 
\ivinst { 12} {R9} {UL14} {} {} 
\ivinst { 13} {L24} {R9} {} {} 
\vskip 2pt
\ligne{\tt state \the\stnumer \global\advance \stnumer by 1\hfill}
\vskip 2pt
\ivinst { 14} {L15} {L16} {L17} {L18} 
\ivinst { 15} {} {L14} {L14} {} 
\ivinst { 16} {L14} {L14} {} {L14} 
\ivinst { 17} {L14} {R} {AR19} {L14} 
\ivinst { 18} {L14} {L14} {} {L14} 
\vskip 2pt
\ligne{\tt state \the\stnumer \global\advance \stnumer by 1\hfill}
\vskip 2pt
\ivinst { 19} {R20} {R21} {R22} {R23} 
\ivinst { 20} {R19} {R19} {R19} {R19} 
\ivinst { 21} {R19} {R19} {} {R19} 
\ivinst { 22} {R19} {R9} {} {CR} 
\ivinst { 23} {} {R19} {CL22} {R19} 
\vskip 2pt
\ligne{\tt state \the\stnumer \global\advance \stnumer by 1\hfill}
\vskip 2pt
\ivinst { 24} {L25} {L26} {L27} {L28} 
\ivinst { 25} {} {L24} {L24} {} 
\ivinst { 26} {L24} {L24} {} {L29} 
\ivinst { 27} {L24} {} {} {} 
\ivinst { 28} {L24} {L24} {} {} 
\vskip 2pt
\ligne{\tt state \the\stnumer \global\advance \stnumer by 1\hfill}
\vskip 2pt
\ivinst { 29} {L30} {} {L31} {L32} 
\ivinst { 30} {UR33} {R} {UR33} {} 
\ivinst { 31} {} {R30} {} {} 
\ivinst { 32} {} {L29} {} {L29} 
\vskip 2pt
\ligne{\tt state \the\stnumer \global\advance \stnumer by 1\hfill}
\vskip 2pt
\ivinst { 33} {R34} {R35} {R36} {R37} 
\ivinst { 34} {R33} {R33} {R33} {R33} 
\ivinst { 35} {R33} {R33} {} {R33} 
\ivinst { 36} {R33} {} {} {AR37} 
\ivinst { 37} {R38} {AL36} {} {R33} 
\vskip 2pt
\ligne{\tt state \the\stnumer \global\advance \stnumer by 1\hfill}
\vskip 2pt
\ivinst { 38} {R39} {R40} {R41} {R42} 
\ivinst { 39} {CL42} {R38} {R38} {R38} 
\ivinst { 40} {R38} {R38} {} {R38} 
\ivinst { 41} {R38} {} {} {L49} 
\ivinst { 42} {UL44} {} {} {L41} 
\ivinst { 43} {} {} {} {L44} 
}
\hfill
\vtop{\leftskip 0pt\parindent 0pt\hsize=\largc\baselineskip 8pt
\ivinst {} A  C  G  U  
\vskip 2pt
\ligne{\tt state \the\stnumer \global\advance \stnumer by 1\hfill} 
\vskip 2pt
\ivinst { 44} {L45} {L46} {L47} {L48} 
\ivinst { 45} {L44} {L44} {L44} {} 
\ivinst { 46} {L44} {L44} {} {} 
\ivinst { 47} {L44} {} {} {} 
\ivinst { 48} {L44} {L44} {} {L29} 
\vskip 2pt
\ligne{\tt state \the\stnumer \global\advance \stnumer by 1\hfill}
\vskip 2pt
\ivinst { 49} {L50} {L51} {L52} {L53} 
\ivinst { 50} {L49} {L49} {L49} {} 
\ivinst { 51} {L49} {L49} {CR54} {} 
\ivinst { 52} {L49} {UR51} {} {} 
\ivinst { 53} {L49} {L49} {} {L49} 
\vskip 2pt
\ligne{\hrulefill}
\vskip 5pt
\ligne{\tt state \the\stnumer \global\advance \stnumer by 1\hfill}
\vskip 2pt
\ivinst { 54} {} {R55} {} {} 
\ivinst { 55} {} {GR56} {} {CL} 
\ivinst { 56} {} {R57} {} {} 
\vskip 2pt
\ligne{\tt state \the\stnumer \global\advance \stnumer by 1\hfill}
\vskip 2pt
\ivinst { 57} {} {R58} {} {} 
\ivinst { 58} {} {GR59} {} {CL} 
\ivinst { 59} {} {R60} {} {} 
\vskip 2pt
\ligne{\tt state \the\stnumer \global\advance \stnumer by 1\hfill}
\vskip 2pt
\ivinst { 60} {} {R61} {} {} 
\ivinst { 61} {R62} {GR} {} {AL} 
\vskip 2pt
\ligne{\tt state \the\stnumer \global\advance \stnumer by 1\hfill}
\vskip 2pt
\ivinst { 62} {} {R63} {} {} 
\ivinst { 63} {} {GL65} {} {CL} 
\ivinst { 64} {} {L65} {} {} 
\vskip 2pt
\ligne{\tt state \the\stnumer \global\advance \stnumer by 1\hfill}
\vskip 2pt
\ivinst { 65} {L66} {L67} {L68} {L69} 
\ivinst { 66} {GR70} {R80} {AR} {} 
\ivinst { 67} {} {UR70} {R} {R66} 
\ivinst { 68} {L65} {} {} {} 
\ivinst { 69} {} {} {L65} {} 
\vskip 2pt
\ligne{\tt state \the\stnumer \global\advance \stnumer by 1\hfill}
\vskip 2pt
\ivinst { 70} {R71} {R72} {R73} {R74} 
\ivinst { 71} {} {GL75} {R70} {L88} 
\ivinst { 72} {} {} {} {AL71} 
\ivinst { 73} {R70} {R70} {} {R70} 
\ivinst { 74} {} {} {} {L71} 
\vskip 2pt
\ligne{\tt state \the\stnumer \global\advance \stnumer by 1\hfill}
\vskip 2pt
\ivinst { 75} {L76} {L77} {L78} {L79} 
\ivinst { 76} {} {UR70} {L75} {} 
\ivinst { 77} {} {} {R76} {} 
\ivinst { 78} {L75} {} {} {} 
\ivinst { 79} {} {} {L65} {} 
\vskip 2pt
\ligne{\tt state \the\stnumer \global\advance \stnumer by 1\hfill}
\vskip 2pt
\ivinst { 80} {R81} {R82} {R83} {R84} 
\ivinst { 81} {} {} {R80} {L85} 
\ivinst { 82} {} {L85} {} {L} 
\ivinst { 83} {R80} {} {} {R80} 
\ivinst { 84} {} {} {} {L81} 
}
\hfill}

\ligne{\hfill
\vtop{\leftskip 0pt\parindent 0pt\hsize=\largc\baselineskip 8pt
\ivinst {} A  C  G  U  
\vskip 2pt
\ligne{\tt state \the\stnumer \global\advance \stnumer by 1\hfill} 
\vskip 2pt
\ivinst { 85} {L86} {} {} {L87} 
\ivinst { 86} {CL87} {} {R} {CL} 
\ivinst { 87} {} {GR86} {L65} {} 
\vskip 2pt
\ligne{\tt state \the\stnumer \global\advance \stnumer by 1\hfill}
\vskip 2pt
\ivinst { 88} {L89} {L90} {L91} {L92} 
\ivinst { 89} {} {R102} {CL88} {CL90} 
\ivinst { 90} {GR97} {} {L88} {R102} 
\ivinst { 91} {GR} {} {AL95} {R89} 
\ivinst { 92} {} {} {R89} {} 
\ivinst { 93} {R94} {} {R95} {R96} 
\ivinst { 94} {} {} {AL90} {} 
\ivinst { 95} {L89} {L90} {L97} {CL90} 
\ivinst { 96} {} {R102} {} {} 
\vskip 2pt
\ligne{\tt state \the\stnumer \global\advance \stnumer by 1\hfill}
\vskip 2pt
\ivinst { 97} {L98} {L99} {L100} {L101} 
\ivinst { 98} {} {R102} {L97} {CL} 
\ivinst { 99} {} {} {AL98} {R98} 
\ivinst {100} {GR99} {} {} {} 
\ivinst {101} {} {} {R98} {} 
\vskip 2pt
\ligne{\hrulefill}
\vskip 5pt
\ligne{\tt state \the\stnumer \global\advance \stnumer by 1\hfill}
\vskip 2pt
\ivinst {102} {} {R103} {R104} {R105} 
\ivinst {103} {UR102} {} {CL115} {R102} 
\ivinst {104} {UL103} {UL} {CL106} {} 
\ivinst {105} {} {R88} {} {L120} 
\vskip 2pt
\ligne{\tt state \the\stnumer \global\advance \stnumer by 1\hfill}
\vskip 2pt
\ivinst {106} {L107} {L108} {L109} {L110} 
\ivinst {107} {} {L106} {CR111} {} 
\ivinst {108} {} {} {L106} {L106} 
\ivinst {109} {} {GR107} {} {} 
\ivinst {110} {} {L106} {} {} 
\vskip 2pt
\ligne{\tt state \the\stnumer \global\advance \stnumer by 1\hfill}
\vskip 2pt
\ivinst {111} {} {R112} {R113} {R114} 
\ivinst {112} {R111} {} {} {} 
\ivinst {113} {R111} {R111} {} {} 
\ivinst {114} {} {R102} {} {} 
\vskip 2pt
\ligne{\tt state \the\stnumer \global\advance \stnumer by 1\hfill}
\vskip 2pt
\ivinst {115} {L116} {L117} {L118} {L119} 
\ivinst {116} {} {L115} {} {} 
\ivinst {117} {} {} {L115} {L115} 
\ivinst {118} {} {R} {AR111} {} 
\ivinst {119} {} {L115} {} {} 
\vskip 2pt
\ligne{\tt state \the\stnumer \global\advance \stnumer by 1\hfill}
\vskip 2pt
\ivinst {120} {} {L121} {} {L122} 
\ivinst {121} {} {R124} {} {R} 
\ivinst {122} {} {R} {} {AL123} 
\ivinst {123} {} {L120} {} {} 
\vskip 2pt
\ligne{\tt state \the\stnumer \global\advance \stnumer by 1\hfill}
\vskip 2pt
\ivinst {124} {R125} {R126} {} {R127} 
\ivinst {125} {GL127} {R124} {R124} {R124} 
\ivinst {126} {R124} {R124} {R128} {R124} 
\ivinst {127} {UR126} {} {} {R124} 
\vskip 2pt
\ligne{\tt state \the\stnumer \global\advance \stnumer by 1\hfill}
\vskip 2pt
\ivinst {128} {R129} {R130} {R131} {R132} 
\ivinst {129} {R128} {R128} {R128} {R128} 
\ivinst {130} {R128} {R128} {} {R128} 
\ivinst {131} {R128} {GL138} {R128} {} 
\ivinst {132} {} {} {R133} {R128} 
\vskip 2pt
\ligne{\hrulefill}
}
\hfill
\vtop{\leftskip 0pt\parindent 0pt\hsize=\largc\baselineskip 8pt
\ivinst {} A  C  G  U  
\vskip 2pt
\ligne{\tt state \the\stnumer \global\advance \stnumer by 1\hfill} 
\vskip 2pt
\ivinst {133} {R134} {R135} {R136} {R137} 
\ivinst {134} {L137} {L169} {L169} {} 
\ivinst {135} {CL131} {L134} {} {L137} 
\ivinst {136} {L134} {R133} {} {R133} 
\ivinst {137} {L169} {GL159} {} {} 
\vskip 2pt
\ligne{\tt state \the\stnumer \global\advance \stnumer by 1\hfill}
\vskip 2pt
\ivinst {138} {L139} {L140} {L141} {L142} 
\ivinst {139} {L138} {L138} {L138} {L138} 
\ivinst {140} {L138} {L138} {L138} {L143} 
\ivinst {141} {L138} {} {L138} {L138} 
\ivinst {142} {L138} {L138} {L138} {L138} 
\vskip 2pt
\ligne{\tt state \the\stnumer \global\advance \stnumer by 1\hfill}
\vskip 2pt
\ivinst {143} {L144} {L145} {L146} {L147} 
\ivinst {144} {} {L143} {L143} {} 
\ivinst {145} {} {AR148} {R} {} 
\ivinst {146} {} {} {L148} {} 
\ivinst {147} {} {} {L143} {} 
\vskip 2pt
\ligne{\tt state \the\stnumer \global\advance \stnumer by 1\hfill}
\vskip 2pt
\ivinst {148} {R149} {R150} {R151} {R152} 
\ivinst {149} {R148} {R148} {R148} {R148} 
\ivinst {150} {R148} {R148} {} {R148} 
\ivinst {151} {R148} {} {} {R148} 
\ivinst {152} {R148} {R148} {R153} {R148} 
\vskip 2pt
\ligne{\tt state \the\stnumer \global\advance \stnumer by 1\hfill}
\vskip 2pt
\ivinst {153} {R154} {R155} {R156} {R157} 
\ivinst {154} {R153} {R153} {R153} {R153} 
\ivinst {155} {R153} {R153} {CR158} {R153} 
\ivinst {156} {R153} {AL155} {R153} {L158} 
\ivinst {157} {R153} {} {R133} {R153} 
\ivinst {158} {R133} {} {CR} {R133} 
\vskip 2pt
\ligne{\tt state \the\stnumer \global\advance \stnumer by 1\hfill}
\vskip 2pt
\ivinst {159} {L160} {L161} {L162} {L163} 
\ivinst {160} {L159} {L159} {L159} {L159} 
\ivinst {161} {L159} {L159} {L159} {L164} 
\ivinst {162} {L159} {} {L159} {L159} 
\ivinst {163} {L159} {L159} {L159} {L159} 
\vskip 2pt
\ligne{\tt state \the\stnumer \global\advance \stnumer by 1\hfill}
\vskip 2pt
\ivinst {164} {L165} {L166} {L167} {L168} 
\ivinst {165} {} {L164} {L164} {} 
\ivinst {166} {} {UR148} {R} {} 
\ivinst {167} {} {} {} {} 
\ivinst {168} {} {} {L164} {} 
\vskip 2pt
\ligne{\tt state \the\stnumer \global\advance \stnumer by 1\hfill}
\vskip 2pt
\ivinst {169} {L170} {L171} {L172} {L173} 
\ivinst {170} {L169} {L169} {L169} {L169} 
\ivinst {171} {L169} {L169} {CR172} {L174} 
\ivinst {172} {L169} {AL171} {L169} {L169} 
\ivinst {173} {L169} {L169} {CL169} {L169} 
\vskip 2pt
\ligne{\tt state \the\stnumer \global\advance \stnumer by 1\hfill}
\vskip 2pt
\ivinst {174} {L175} {L176} {L177} {L178} 
\ivinst {175} {} {L174} {L174} {} 
\ivinst {176} {} {} {L174} {} 
\ivinst {177} {} {R} {R179} {} 
\ivinst {178} {} {L174} {L174} {} 
\vskip 2pt
\ligne{\tt state \the\stnumer \global\advance \stnumer by 1\hfill}
\vskip 2pt
\ivinst {179} {} {R180} {R181} {R182} 
\ivinst {180} {R179} {} {AR182} {} 
\ivinst {181} {R179} {R179} {} {CL180} 
\ivinst {182} {} {R183} {} {} 
\vskip 2pt
\ligne{\hrulefill}
}
\hfill}

\ligne{\hfill
\vtop{\leftskip 0pt\parindent 0pt\hsize=\largc\baselineskip 8pt
\ivinst {} A  C  G  U  
\vskip 2pt
\ligne{\tt state \the\stnumer \global\advance \stnumer by 1\hfill} 
\vskip 2pt
\ivinst {183} {R184} {R184} {R185} {R186} 
\ivinst {184} {R183} {} {} {} 
\ivinst {185} {R183} {AR183} {R183} {R183} 
\ivinst {186} {} {R187} {} {} 
\vskip 2pt
\ligne{\tt state \the\stnumer \global\advance \stnumer by 1\hfill}
\vskip 2pt
\ivinst {187} {} {R188} {R189} {} 
\ivinst {188} {L} {GL191} {} {} 
\ivinst {189} {R187} {} {} {} 
\ivinst {190} {} {} {} {} 
\vskip 2pt
\ligne{\tt state \the\stnumer \global\advance \stnumer by 1\hfill}
\vskip 2pt
\ivinst {191} {L192} {L193} {L194} {L195} 
\ivinst {192} {} {L191} {L191} {} 
\ivinst {193} {R194} {L191} {L191} {L191} 
\ivinst {194} {L191} {R196} {L191} {} 
\ivinst {195} {} {L191} {L191} {} 
\vskip 2pt
\ligne{\tt state \the\stnumer \global\advance \stnumer by 1\hfill}
\vskip 2pt
\ivinst {196} {R197} {R198} {R199} {R200} 
\ivinst {197} {L393} {} {R196} {L} 
\ivinst {198} {CR183} {R196} {AR200} {L201} 
\ivinst {199} {GR183} {} {R196} {GL198} 
\ivinst {200} {} {L198} {R183} {} 
\vskip 2pt
\ligne{\tt state \the\stnumer \global\advance \stnumer by 1\hfill}
\vskip 2pt
\ivinst {201} {} {L202} {L203} {L204} 
\ivinst {202} {GR203} {R204} {UL204} {} 
\ivinst {203} {GR202} {R206} {R205} {} 
\ivinst {204} {GR203} {AL205} {L201} {} 
\ivinst {205} {} {L201} {CL204} {} 
\vskip 2pt
\ligne{\tt state \the\stnumer \global\advance \stnumer by 1\hfill}
\vskip 2pt
\ivinst {206} {} {R207} {R208} {R209} 
\ivinst {207} {R206} {R183} {CR209} {} 
\ivinst {208} {L207} {R206} {AR207} {CL} 
\ivinst {209} {R210} {R206} {} {} 
\vskip 2pt
\ligne{\tt state \the\stnumer \global\advance \stnumer by 1\hfill}
\vskip 2pt
\ivinst {210} {R211} {R212} {R213} {R214} 
\ivinst {211} {R210} {R210} {R210} {R210} 
\ivinst {212} {R210} {R210} {} {R210} 
\ivinst {213} {R210} {} {} {} 
\ivinst {214} {R374} {} {R215} {R210} 
\vskip 2pt
\ligne{\tt state \the\stnumer \global\advance \stnumer by 1\hfill}
\vskip 2pt
\ivinst {215} {R216} {R217} {R218} {R219} 
\ivinst {216} {} {R215} {R215} {R215} 
\ivinst {217} {R215} {L218} {} {R215} 
\ivinst {218} {R215} {UL220} {} {} 
\ivinst {219} {} {} {} {} 
\vskip 2pt
\ligne{\tt state \the\stnumer \global\advance \stnumer by 1\hfill}
\vskip 2pt
\ivinst {220} {L221} {L222} {L223} {L224} 
\ivinst {221} {L220} {L220} {L220} {} 
\ivinst {222} {L220} {L220} {} {} 
\ivinst {223} {L220} {} {} {L220} 
\ivinst {224} {L220} {L220} {} {L225} 
\vskip 2pt
\ligne{\tt state \the\stnumer \global\advance \stnumer by 1\hfill}
\vskip 2pt
\ivinst {225} {L226} {L227} {} {} 
\ivinst {226} {R228} {L225} {CR} {} 
\ivinst {227} {} {R233} {} {R} 
\vskip 2pt
\ligne{\tt state \the\stnumer \global\advance \stnumer by 1\hfill} 
\vskip 2pt
\ivinst {228} {R229} {R230} {R231} {R232} 
\ivinst {229} {R228} {R228} {R228} {R228} 
\ivinst {230} {R228} {R228} {} {R228} 
\ivinst {231} {R228} {R215} {} {CR} 
\ivinst {232} {} {L231} {R228} {R228} 
\vskip 2pt
\ligne{\hrulefill}
}
\hfill
\vtop{\leftskip 0pt\parindent 0pt\hsize=\largc\baselineskip 8pt
\ivinst {} A  C  G  U  
\vskip 2pt
\ligne{\tt state \the\stnumer \global\advance \stnumer by 1\hfill}
\vskip 2pt
\ivinst {233} {R234} {R235} {R236} {R237} 
\ivinst {234} {R233} {R233} {R233} {R233} 
\ivinst {235} {R233} {R233} {} {R233} 
\ivinst {236} {R233} {} {} {} 
\ivinst {237} {} {GR238} {R233} {R233} 
\vskip 2pt
\ligne{\tt state \the\stnumer \global\advance \stnumer by 1\hfill}
\vskip 2pt
\ivinst {238} {R239} {R240} {R241} {} 
\ivinst {239} {} {GR242} {CR} {} 
\ivinst {240} {CL239} {R238} {L282} {L239} 
\ivinst {241} {L240} {R238} {} {R238} 
\ivinst {242} {} {R243} {} {R263} 
\vskip 2pt
\ligne{\tt state \the\stnumer \global\advance \stnumer by 1\hfill}
\vskip 2pt
\ivinst {243} {R244} {R245} {R246} {R247} 
\ivinst {244} {R243} {R243} {R243} {R243} 
\ivinst {245} {R243} {R243} {} {R243} 
\ivinst {246} {R243} {} {R243} {} 
\ivinst {247} {R243} {} {R248} {R243} 
\vskip 2pt
\ligne{\tt state \the\stnumer \global\advance \stnumer by 1\hfill}
\vskip 2pt
\ivinst {248} {R249} {R250} {R251} {R252} 
\ivinst {249} {} {GL253} {L253} {} 
\ivinst {250} {CL249} {} {} {CL249} 
\ivinst {251} {CL249} {R248} {} {R248} 
\ivinst {252} {} {} {} {} 
\vskip 2pt
\ligne{\tt state \the\stnumer \global\advance \stnumer by 1\hfill}
\vskip 2pt
\ivinst {253} {} {L254} {L255} {L256} 
\ivinst {254} {} {} {L253} {} 
\ivinst {255} {} {} {} {L257} 
\ivinst {256} {} {} {L253} {} 
\vskip 2pt
\ligne{\tt state \the\stnumer \global\advance \stnumer by 1\hfill}
\vskip 2pt
\ivinst {257} {L258} {L259} {L260} {L261} 
\ivinst {258} {L257} {L257} {L257} {} 
\ivinst {259} {L257} {L257} {R260} {R238} 
\ivinst {260} {L257} {R238} {L257} {R262} 
\ivinst {261} {L257} {L257} {R260} {L257} 
\ivinst {262} {} {} {R238} {} 
\vskip 2pt
\ligne{\tt state \the\stnumer \global\advance \stnumer by 1\hfill}
\vskip 2pt
\ivinst {263} {R264} {R265} {R266} {R267} 
\ivinst {264} {R263} {R263} {R263} {R263} 
\ivinst {265} {R263} {R263} {} {R263} 
\ivinst {266} {R263} {} {R263} {} 
\ivinst {267} {} {} {R268} {R263} 
\vskip 2pt
\ligne{\tt state \the\stnumer \global\advance \stnumer by 1\hfill}
\vskip 2pt
\ivinst {268} {} {R269} {R270} {} 
\ivinst {269} {UL} {GL271} {} {L} 
\ivinst {270} {UL} {R268} {L271} {R268} 
\vskip 2pt
\ligne{\tt state \the\stnumer \global\advance \stnumer by 1\hfill}
\vskip 2pt
\ivinst {271} {L272} {L273} {L274} {L275} 
\ivinst {272} {} {} {L271} {} 
\ivinst {273} {} {} {L271} {} 
\ivinst {274} {} {} {} {L276} 
\ivinst {275} {} {} {L271} {L271} 
\vskip 2pt
\ligne{\tt state \the\stnumer \global\advance \stnumer by 1\hfill} 
\vskip 2pt
\ivinst {276} {L277} {L278} {L279} {L280} 
\ivinst {277} {L276} {L276} {L276} {} 
\ivinst {278} {L276} {L276} {L276} {L276} 
\ivinst {279} {L276} {} {L276} {R281} 
\ivinst {280} {L276} {L276} {L276} {L276} 
\ivinst {281} {} {} {R238} {} 
\vskip 2pt
\ligne{\hrulefill}
}
\hfill}

\ligne{\hfill
\vtop{\leftskip 0pt\parindent 0pt\hsize=\largc\baselineskip 8pt
\ivinst {} A  C  G  U  
\vskip 2pt
\ligne{\tt state \the\stnumer \global\advance \stnumer by 1\hfill}
\vskip 2pt
\ivinst {282} {L283} {L284} {L285} {L286} 
\ivinst {283} {} {L282} {R287} {} 
\ivinst {284} {} {AL286} {CR} {} 
\ivinst {285} {} {} {} {R283} 
\ivinst {286} {} {L282} {CL282} {} 
\vskip 2pt
\ligne{\tt state \the\stnumer \global\advance \stnumer by 1\hfill}
\vskip 2pt
\ivinst {287} {R288} {R289} {R290} {R291} 
\ivinst {288} {R287} {R287} {R287} {R287} 
\ivinst {289} {R287} {R287} {} {R287} 
\ivinst {290} {R287} {} {R287} {} 
\ivinst {291} {} {} {R292} {R287} 
\vskip 2pt
\ligne{\tt state \the\stnumer \global\advance \stnumer by 1\hfill}
\vskip 2pt
\ivinst {292} {} {R293} {R294} {R295} 
\ivinst {293} {L} {GR295} {CR295} {AL} 
\ivinst {294} {L} {AL293} {L296} {L293} 
\ivinst {295} {R292} {} {} {R292} 
\vskip 2pt
\ligne{\tt state \the\stnumer \global\advance \stnumer by 1\hfill}
\vskip 2pt
\ivinst {296} {L297} {L298} {L299} {L300} 
\ivinst {297} {L296} {L296} {L296} {} 
\ivinst {298} {L296} {L296} {L296} {} 
\ivinst {299} {L296} {} {L296} {L301} 
\ivinst {300} {L296} {L296} {L296} {L296} 
\vskip 2pt
\ligne{\tt state \the\stnumer \global\advance \stnumer by 1\hfill}
\vskip 2pt
\ivinst {301} {L302} {L303} {L304} {L305} 
\ivinst {302} {L301} {L301} {L301} {} 
\ivinst {303} {L301} {L301} {L301} {} 
\ivinst {304} {L301} {} {L301} {R306} 
\ivinst {305} {L301} {L301} {L301} {L301} 
\ivinst {306} {} {} {R307} {} 
\vskip 2pt
\ligne{\tt state \the\stnumer \global\advance \stnumer by 1\hfill}
\vskip 2pt
\ivinst {307} {R308} {R309} {R310} {R311} 
\ivinst {308} {GR309} {GL} {CL311} {} 
\ivinst {309} {R307} {R312} {R312} {R307} 
\ivinst {310} {R307} {R322} {} {} 
\ivinst {311} {UR310} {} {} {} 
\vskip 2pt
\ligne{\tt state \the\stnumer \global\advance \stnumer by 1\hfill}
\vskip 2pt
\ivinst {312} {R313} {R314} {R315} {R316} 
\ivinst {313} {R312} {R312} {R312} {R312} 
\ivinst {314} {R312} {R312} {} {R312} 
\ivinst {315} {R312} {} {R312} {GL317} 
\ivinst {316} {} {} {L315} {R312} 
\vskip 2pt
\ligne{\tt state \the\stnumer \global\advance \stnumer by 1\hfill}
\vskip 2pt
\ivinst {317} {L318} {L319} {L320} {L321} 
\ivinst {318} {} {L317} {L317} {} 
\ivinst {319} {} {} {} {} 
\ivinst {320} {} {} {UL353} {} 
\ivinst {321} {} {L317} {} {} 
\vskip 2pt
\ligne{\tt state \the\stnumer \global\advance \stnumer by 1\hfill}
\vskip 2pt
\ivinst {322} {R323} {R324} {R325} {R326} 
\ivinst {323} {R322} {R322} {R322} {R322} 
\ivinst {324} {R322} {R322} {R327} {R322} 
\ivinst {325} {R322} {} {R322} {GR324} 
\ivinst {326} {} {} {L325} {R322} 
\vskip 2pt
\ligne{\tt state \the\stnumer \global\advance \stnumer by 1\hfill} 
\vskip 2pt
\ivinst {327} {R328} {R329} {R330} {R331} 
\ivinst {328} {} {R327} {R327} {R327} 
\ivinst {329} {R327} {GL332} {L} {R327} 
\ivinst {330} {R327} {} {L331} {} 
\ivinst {331} {} {} {UL353} {} 
\vskip 2pt
\ligne{\hrulefill}
}
\hfill
\vtop{\leftskip 0pt\parindent 0pt\hsize=\largc\baselineskip 8pt
\ivinst {} A  C  G  U  
\vskip 2pt
\ligne{\tt state \the\stnumer \global\advance \stnumer by 1\hfill}
\vskip 2pt
\ivinst {332} {L333} {} {} {} 
\ivinst {333} {CR334} {} {CR} {} 
\vskip 2pt
\ligne{\tt state \the\stnumer \global\advance \stnumer by 1\hfill}
\vskip 2pt
\ivinst {334} {R335} {R336} {R337} {} 
\ivinst {335} {} {GL338} {R334} {} 
\ivinst {336} {} {R334} {AL335} {} 
\ivinst {337} {R334} {} {R334} {R334} 
\vskip 2pt
\ligne{\tt state \the\stnumer \global\advance \stnumer by 1\hfill}
\vskip 2pt
\ivinst {338} {L339} {} {L340} {} 
\ivinst {339} {} {} {L338} {} 
\ivinst {340} {} {} {L341} {} 
\vskip 2pt
\ligne{\tt state \the\stnumer \global\advance \stnumer by 1\hfill}
\vskip 2pt
\ivinst {341} {L342} {L343} {L344} {L345} 
\ivinst {342} {UR334} {GR} {CR343} {GR334} 
\ivinst {343} {CR334} {L341} {R346} {} 
\ivinst {344} {L341} {} {R343} {} 
\ivinst {345} {} {AR342} {L341} {} 
\vskip 2pt
\ligne{\tt state \the\stnumer \global\advance \stnumer by 1\hfill}
\vskip 2pt
\ivinst {346} {R347} {R348} {R349} {R350} 
\ivinst {347} {CR} {GR348} {UL} {R346} 
\ivinst {348} {R346} {AL347} {UR350} {R346} 
\ivinst {349} {UL351} {} {L348} {AL352} 
\ivinst {350} {UL} {} {R346} {} 
\ivinst {351} {} {} {CL353} {} 
\ivinst {352} {} {} {CR348} {} 
\vskip 2pt
\ligne{\tt state \the\stnumer \global\advance \stnumer by 1\hfill}
\vskip 2pt
\ivinst {353} {L354} {L355} {L356} {L357} 
\ivinst {354} {} {L353} {L353} {} 
\ivinst {355} {} {L353} {} {} 
\ivinst {356} {} {} {L353} {L353} 
\ivinst {357} {} {L353} {} {L358} 
\vskip 2pt
\ligne{\tt state \the\stnumer \global\advance \stnumer by 1\hfill}
\vskip 2pt
\ivinst {358} {L359} {L360} {L361} {L362} 
\ivinst {359} {L358} {L358} {L358} {AR363} 
\ivinst {360} {L358} {L358} {} {L358} 
\ivinst {361} {L358} {} {L358} {L358} 
\ivinst {362} {L358} {L358} {} {R359} 
\vskip 2pt
\ligne{\tt state \the\stnumer \global\advance \stnumer by 1\hfill}
\vskip 2pt
\ivinst {363} {R364} {R365} {R366} {R367} 
\ivinst {364} {R363} {R363} {R363} {R363} 
\ivinst {365} {R363} {R363} {} {R363} 
\ivinst {366} {R363} {} {} {AR367} 
\ivinst {367} {R368} {} {AL366} {} 
\vskip 2pt
\ligne{\tt state \the\stnumer \global\advance \stnumer by 1\hfill}
\vskip 2pt
\ivinst {368} {R369} {R370} {R371} {R372} 
\ivinst {369} {R368} {R368} {R368} {R368} 
\ivinst {370} {R368} {R368} {AR371} {R368} 
\ivinst {371} {R368} {R133} {CL370} {AR373} 
\ivinst {372} {} {GL371} {R368} {} 
\ivinst {373} {} {} {R133} {} 
\vskip 2pt
\ligne{\hrulefill}
\vskip 5pt
\ligne{\tt state \the\stnumer \global\advance \stnumer by 1\hfill} 
\vskip 2pt
\ivinst {374} {R375} {R376} {R377} {} 
\ivinst {375} {CL378} {R374} {R374} {R374} 
\ivinst {376} {R374} {R374} {} {R374} 
\ivinst {377} {R374} {} {} {} 
\ivinst {378} {UL393} {} {} {} 
}
\hfill}

\ligne{\hfill
\vtop{\leftskip 0pt\parindent 0pt\hsize=\largc\baselineskip 8pt
\ivinst {} A  C  G  U  
\vskip 2pt
\ligne{\tt state \the\stnumer \global\advance \stnumer by 1\hfill}
\vskip 2pt
\ivinst {379} {} {R380} {R381} {R382} 
\ivinst {380} {R379} {} {} {} 
\ivinst {381} {R379} {} {} {} 
\ivinst {382} {R383} {} {} {} 
\vskip 2pt
\ligne{\tt state \the\stnumer \global\advance \stnumer by 1\hfill}
\vskip 2pt
\ivinst {383} {R384} {R385} {R386} {R387} 
\ivinst {384} {R383} {R383} {R383} {R383} 
\ivinst {385} {R383} {R383} {} {R383} 
\ivinst {386} {R383} {} {} {AR387} 
\ivinst {387} {R388} {AL386} {} {} 
\vskip 2pt
\ligne{\tt state \the\stnumer \global\advance \stnumer by 1\hfill}
\vskip 2pt
\ivinst {388} {R389} {R390} {R391} {R392} 
\ivinst {389} {CL392} {R388} {R388} {R388} 
\ivinst {390} {R388} {R388} {} {R388} 
\ivinst {391} {R388} {} {} {CR390} 
\ivinst {392} {UL393} {} {CL391} {} 
\vskip 2pt
\ligne{\tt state \the\stnumer \global\advance \stnumer by 1\hfill}
\vskip 2pt
\ivinst {393} {L394} {L395} {L396} {L397} 
\ivinst {394} {L393} {L393} {L393} {L393} 
\ivinst {395} {L393} {L393} {L393} {L393} 
\ivinst {396} {L393} {R} {R398} {L393} 
\ivinst {397} {L393} {L393} {} {} 
}
\hfill
\vtop{\leftskip 0pt\parindent 0pt\hsize=\largc\baselineskip 8pt
\ivinst {} A  C  G  U  
\vskip 2pt
\ligne{\tt state \the\stnumer \global\advance \stnumer by 1\hfill}
\vskip 2pt
\ivinst {398} {} {R399} {R400} {R401} 
\ivinst {399} {R398} {} {CR402} {} 
\ivinst {400} {L399} {R398} {} {} 
\ivinst {401} {UR403} {R398} {} {} 
\ivinst {402} {R379} {} {} {} 
\vskip 2pt
\ligne{\tt state \the\stnumer \global\advance \stnumer by 1\hfill}
\vskip 2pt
\ivinst {403} {R404} {R405} {R406} {R407} 
\ivinst {404} {R403} {R403} {R403} {R403} 
\ivinst {405} {R403} {R403} {} {R403} 
\ivinst {406} {R403} {} {} {CR405} 
\ivinst {407} {R403} {GR403} {CL406} {L408} 
\ivinst {408} {} {} {} {L409} 
\vskip 2pt
\ligne{\tt state \the\stnumer \global\advance \stnumer by 1\hfill}
\vskip 2pt
\ivinst {409} {L410} {L411} {L412} {L413} 
\ivinst {410} {L409} {L409} {L409} {} 
\ivinst {411} {L409} {L409} {} {} 
\ivinst {412} {L409} {} {R128} {R} 
\ivinst {413} {L409} {L409} {} {} 
}
\hfill}

\end{document}